\documentclass[floatfix,aps,amsmath,prd,twocolumn,notitlepage,eqsecnum,showpacs,nofootinbib,aasmacros,superscriptaddress]{revtex4-1}
\usepackage{bm, graphics, graphicx, epsfig, soul, latexsym, hyperref ,multirow}
\usepackage{hyperref}
\usepackage{comment}
\usepackage{subfigure}
\usepackage{graphicx,color}
\usepackage{dcolumn}
\usepackage{bm,amsmath, amssymb,}
\usepackage{mathrsfs}
\usepackage{bigints}
\usepackage{float}
\usepackage{multirow}
\usepackage{footmisc}
\usepackage{placeins}
\usepackage[normalem]{ulem}

\usepackage{amsmath}

\newcommand{\be}{\begin{equation}}
\newcommand{\ee}{\end{equation}}
\begin{document}
\title{Systematic bias on the inspiral-merger-ringdown consistency test due to neglect of orbital eccentricity }
\author{Sajad A.~Bhat}
\email{sabhat@cmi.ac.in}
\affiliation{Chennai Mathematical Institute, Siruseri 603103, India}
\author{Pankaj Saini}
\email{pankajsaini@cmi.ac.in}
\affiliation{Chennai Mathematical Institute, Siruseri 603103, India}
\author{Marc Favata}
\email{marc.favata@montclair.edu}
\affiliation{Department of Physics \& Astronomy, Montclair State University,
1 Normal Avenue, Montclair, New Jersey 07043, USA}
\author{K.~G.~Arun}
\email{kgarun@cmi.ac.in}
\affiliation{Chennai Mathematical Institute, Siruseri 603103, India}
\date{\today}
 \begin{abstract}
The inspiral-merger-ringdown (IMR) consistency test checks the consistency of the final mass and final spin of a binary black hole merger remnant, independently inferred via the inspiral and merger-ringdown parts of the waveform. As binaries are expected to be nearly circularized when entering the frequency band of ground-based detectors, tests of general relativity (GR) currently employ quasicircular waveforms. We quantify the effect of residual orbital eccentricity on the IMR consistency test. We find that eccentricity causes a significant systematic bias in the inferred final mass and spin of the remnant black hole at an orbital eccentricity (defined at $10$ Hz) of $e_0 \gtrsim 0.1$ in the LIGO band (for a total binary mass in the range $65$-$200 \,M_{\odot}$). For binary black holes observed by Cosmic Explorer (CE), the systematic bias becomes significant for  $e_0 \gtrsim 0.015$  (for $200$-$600 \,M_{\odot}$ systems). 
This eccentricity-induced bias on the final mass and spin leads to an apparent inconsistency in the IMR consistency test, manifesting as a false violation of GR. Hence, eccentric corrections to waveform models are important for constructing a robust test of GR, especially for third-generation detectors. We also estimate the eccentric corrections to the relationship between the inspiral parameters and the final mass and final spin; they are shown to be quite small.
\end{abstract}
\maketitle
\section{Introduction}
With the increasing number of gravitational-wave (GW) detections by LIGO/Virgo\cite{abbott2019gwtc1,abbott2021gwtc2,abbott2021gwtc3,LIGOScientific:2014pky,VIRGO:2014yos}, general relativity (GR) has been subjected to a battery of tests in the strong field and nonlinear regime of the theory. No statistically significant evidence of physics beyond GR was found in any of these tests \cite{holberg2010sirius,dyson1920ix,Weisberg:2004hi,hees2017testing,voisin2020improved,GW150914TGR:2016lio,abbott2016gw151226,GWTC1TGR:2019fpa,GW170817TGR:2018dkp,GWTC2:2020tif,GWTC3:2021sio,Kramer:2021jcw}. However, efforts in the near and long term aim to test GR with increasing levels of precision. To confidently claim a deviation from GR, the waveform templates that are compared with the data should be free of systematic biases. These might be due to unmodeled physical effects like eccentricity, spin precession, higher modes, or high post-Newtonian-order terms~\cite{Favata:2013rwa,PhysRevD.98.024019}. If these effects are not included at a level consistent with the precision of the detector, systematic biases may lead to a false indication of a GR violation.

One of the many proposed tests of GR using gravitational waves is the {\it inspiral-merger-ringdown} (IMR) {\it consistency test} \cite{Ghosh:2016qgn,Hughes:2004vw}. The IMR consistency test considers the mass and spin of the remnant black hole (BH) obtained via two independent estimates. The individual masses and spins of the component black holes are first inferred from the inspiral (low-frequency) part of the GW signal. Using numerical relativity fits \cite{Healy:2016lce,Hofmann:2016yih,Jimenez-Forteza:2016oae}, these component masses and spins are mapped to the final mass and final spin of the remnant black hole formed after the binary components merge. The final mass and final spin of the remnant black hole can also be independently estimated from the merger-ringdown (high-frequency) part of the signal. If the signal is described by the dynamics of binary black holes in GR, the remnant final mass and spin inferred via these two approaches should be consistent with each other \cite{Ghosh:2017gfp,Carson:2019rda,Nakano:2021bbw}. Lack of consistency indicates a potential GR violation.

Measurements of the final mass and final spin of the remnant black hole using the merger-ringdown signal are poorly constrained with current detectors---due to the low signal-to-noise ratios (SNRs) associated with that part of the signal \cite{GW150914TGR:2016lio,GWTC1TGR:2019fpa,GWTC2:2020tif,GWTC3:2021sio}. This limits the overall precision of the IMR consistency test. The combined posterior on the fractional mass and spin deviation parameters of GWTC-3 events are consistent with GR predictions (see Sec.~IVB of \cite{GWTC3:2021sio}).

It is likely that deviations from GR are too small to be detected by the LIGO/Virgo/Kagra network. The future space-based detector LISA \cite{Babak:2017tow} or third-generation (3G) ground-based detectors \cite{Reitze:2019iox,LIGOScientific:2016wof,Punturo_2010}---which may achieve levels of precision $\sim 10$ to $100$ times greater---are more likely to detect GR deviations. These detectors will attain higher precision measurements of the final mass and final spin inferred from both the inspiral and the merger-ringdown.
In the case of the IMR-consistency test, the inverse of the area enclosed by the error ellipses in the final mass-final spin plane represents a parameter (called the \emph{resolving power}) that characterizes a detector's ability to distinguish between GR and non-GR effects \cite{Carson:2019rda}. Smaller areas or larger resolving powers imply a greater chance of detecting potential GR violations. In terms of this parameter, a 3G detector like Cosmic Explorer (CE) \cite{Reitze:2019iox} results in a factor of $\sim 1000$ improvement in the precision of the IMR consistency test \cite{Carson:2019rda} relative to LIGO's first observing run (O1) \cite{GWTC1TGR:2019fpa}. An additional factor of $\sim 10$ improvement is achieved by combining CE and LISA observations of a given source (i.e., multibanding) \cite{Carson:2019rda,Carson:2020cqb}. Hence, even a small systematic error may contaminate the IMR consistency test.

Since the emission of GWs in a bound binary leads to the rapid decay of orbital eccentricity, it is expected that binaries formed long before they merge will have nearly circularized when they enter the frequency band of ground-based detectors \cite{Peters}. For example, consider a binary with a moderate eccentricity of $\sim 0.2$ when emitting a GW frequency of $0.1$ Hz. When it enters the LIGO frequency band at $10$ Hz, its orbital eccentricity reduces to $\sim$ $10^{-3}$. Hence, LIGO/Virgo analyses primarily employ quasicircular waveforms for parameter estimation of the observed events. 

However, the dynamical formation of compact binaries \cite{Mapelli:2021taw,Postnov:2014tza} in dense environments such as globular clusters and nuclear star clusters may lead to the formation of highly-eccentric binaries. When these eccentric sources enter the frequency band of GW detectors, they still possess some residual eccentricity \cite{Antonini:2012ad,Antonini:2013tea,OLeary:2008myb,Wen:2002km}. The detection of eccentric signals in ground-based detectors using quasicircular waveforms has been studied in detail \cite{Martel:1999tm,Mandel:2007hi,Tessmer:2007jg,Cokelaer:2009hj,Brown:2009ng,Huerta:2013qb,Sun:2015bva,LIGOScientific:2019dag,Ramos-Buades:2020eju}. Those studies found that circular templates are sufficient for detection if eccentricities are $\lesssim 0.02$--$0.15$. Moreover, using quasicircular waveforms to analyze an eccentric source will introduce a systematic bias on the inferred parameters \cite{Favata:2013rwa,Favata:2021vhw}. The size of this parameter bias $\Delta^{\rm sys} \theta_a$ depends (approximately) on the square of the eccentricity $e_0$ near the low-frequency band of the GW detector [$\Delta^{\rm sys} \theta_a \sim O(e_0^2)$]. (Throughout this paper we define $e_0$ to be the orbital eccentricity at a reference GW frequency of $f_0=10\,{\rm Hz}$.) Eccentricity-induced errors may already be biasing the parameter estimation of detected sources that are potentially eccentric \cite{Romero-Shaw:2020thy,Gayathri:2020coq,Romero-Shaw:2021ual,OShea:2021ugg}. 
In addition to biasing parameters, residual eccentricity also has the potential to bias tests of GR. All the current tests presume circularity in the waveform models used in the analysis. In Ref.~\cite{Saini:2022igm} we studied the effect of neglecting the orbital eccentricity on parametrized tests of GR. That test considers only the inspiral phase of the binary evolution in the regime where a post-Newtonian (PN) expansion of the GW phase is valid. The GR values of the coefficients of the PN phase expansion are replaced with new coefficients that are perturbed from their GR values. The parametrized test then attempts to constrain the size of those perturbations (with zero perturbation corresponding to GR). In \cite{Saini:2022igm} we found that the systematic bias on parametrized deviation coefficients becomes significant at $e_0 \sim 0.04$ at $10$ Hz in the LIGO band and at $e_0 \sim 0.005$ at $10$ Hz in the CE band.

Here we perform an analogous study, considering the systematic bias that eccentricity induces on the IMR consistency test.
We focus on the bias in the binary masses and spins that accumulates over many GW cycles during the inspiral. That bias then propagates into the numerical relativity fitting formulas that predict the final mass and spin of the remnant BH, potentially exceeding (if $e_0$ is sufficiently large) the statistical errors on those parameters. When compared with the final mass and spin inferred from the merger-ringdown, this may lead to an inconsistency that mimics a GR violation.
Moreover, the effect of systematic waveform errors may grow for a large catalog of GW events \cite{Moore:2021eok}. 

Section~\ref{sec:formalism} of this paper briefly discusses our waveform model and the methods for calculating statistical and systematic errors. (Additional details about our waveform model are provided in Appendix~\ref{App:waveform}.) Section~\ref{methods} explains how we propagate statistical and systematic errors in the component masses and spins to determine the errors in the final mass and spin (including the relevant error ellipses). Our results are discussed in Sec.~\ref{results}, followed by our conclusions in Sec.~\ref{conclusions}. Appendix~\ref{NRfitecc} provides a quasi-Newtonian derivation of the eccentric corrections to the final mass and final spin formulas. Throughout the paper we use geometric units ($G=c=1$).

\section{\label{sec:formalism}Waveform and Parameter estimation}
We use the sky-averaged {\tt IMRPhenomD} waveform model developed in Refs.~\cite{IMRD,IMRD1}. This is a frequency-domain phenomenological model parametrized in terms of the total binary  mass $M=m_1+m_2$ (for binary component masses $m_1$ and $m_2$), the symmetric mass ratio $\eta=m_1 m_2/(m_1+m_2)^2$, the dimensionless spin parameters of the two BHs ($\chi_1,\chi_2)$, the time and phase of coalescence ($t_c$, $\phi_c$), and the luminosity distance to the source $D_L$. The cosmological redshift is accounted for by replacing $M\rightarrow (1+z) M$, where $z$ is the source redshift. The $M$ used here and throughout refers to the source frame binary mass. We adopt the luminosity distance/redshift relation for a flat universe \cite{hogg},
\begin{equation}
\label{eq:dLz}
D_L(z) = \frac{c}{H_0} (1+z) \int_0^z \frac{dz'}{\sqrt{\Omega_M (1+z')^3 + \Omega_{\Lambda}}},
\end{equation} 
with the following cosmological parameters~\cite{Planck:2015fie}:
{$H_{0}=67.90 \,(\rm{km/s})/\rm{Mpc}$}, $\Omega_{M}=0.3065$, and $\Omega_{\Lambda}=0.6935$.

The {\tt IMRPhenomD} waveform includes only the dominant $(l,m)=(2,\pm 2)$ spin-weighted spherical harmonic modes and assumes that the BH spins are aligned or antialigned with the orbital angular momentum (i.e., are nonprecessing). Reference \cite{IMRD1} provides the amplitude and phase of the Fourier transform of the $h_{2,2}$ waveform mode,
\be
\tilde{h}_{2,2}(f) = A_{\rm IMR}(f) e^{-i \phi_{\rm IMR}(f)},
\ee
where $f$ is the GW frequency. Our parameter estimation formalism below depends on the Fourier transform of the detector response $h(t)=h_{+} F_{+} + h_{\times} F_{\times}$ written in the form $\tilde{h} = {\mathcal A}(f) e^{i\Psi(f)}$. Here $h_{+,\times}$ are the GW polarizations, and $F_{+,\times}$ are the detector antenna response functions. The relationships between ${\mathcal A}(f)$ and $A_{\rm IMR}(f)$ and $\Psi(f)$ and $\phi_{\rm IMR}(f)$ for {\tt IMRPhenomD} are provided in Appendix \ref{App:waveform}.

To incorporate the effects of eccentricity into the {\tt IMRPhenomD} waveform,  we implicitly modify the inspiral phase by adding a leading-order [i.e., $O(e_0^2)$] eccentricity correction. When the waveform Fourier transform is expressed in the form ${\tilde h}(f) =  {\mathcal A}(f) e^{i\Psi(f)}$, this correction shifts the phase $\Psi \rightarrow \Psi + \Delta \Psi$, where $\Delta \Psi$ is the 3PN order eccentric correction computed in Eq.~(6.26) of \cite{Moore:2016qxz}. Since we work in the regime where the eccentricity at the reference frequency is small ($e_0 \lesssim 0.2$), and because the instantaneous eccentricity further decays during the remainder of the inspiral, we ignore any (small) eccentric correction to the merger-ringdown part of the waveform. In our approximation the systematic errors depend only on the phase difference induced by eccentricity corrections to the waveform. Since the Cutler and Vallisneri systematic error formalism \cite{Cutler_Vallisneri:2007mi} (discussed further below) allows us to decouple these eccentric corrections from the rest of the waveform [see, e.g., Eq.~(\ref{systematic errors_1})], a direct modification of the {\tt IMRPhenomD} waveform is not needed in our analysis.  We also ignore eccentricity effects on the mapping between the inspiral parameters and the final mass $M_f$ and final spin $\chi_{f}$ of the merger remnant. Appendix \ref{NRfitecc} computes the leading-order eccentric correction to that mapping and demonstrates that it is small. 

To carry out the IMR consistency test, the signal is divided into two parts: inspiral and merger-ringdown. Both parts of the signal should have sufficient signal-to-noise ratios (SNRs) to allow for precise parameter estimation. The choice of transition frequency from inspiral to merger-ringdown is not unique. However, small variations of the inspiral cut-off frequency do not have a significant impact on the test \cite{Ghosh:2016qgn,Ghosh:2017gfp}. We adopt the choice of the LIGO-Virgo-Kagra (LVK) collaboration \cite{GWTC1TGR:2019fpa,GWTC2:2020tif, GWTC3:2021sio} and choose this transition frequency to be the GW frequency ($f_{\rm ISCO}$) corresponding to the test-particle innermost stable circular orbit (ISCO) \cite{1972ApJ...178..347B} of the remnant Kerr black hole. (This is described further below.)

To calculate the statistical measurement precision, we apply the Fisher information matrix formalism to the {\tt IMRPhenomD} waveform model. This produces the covariance matrix and the $1 \sigma$ width of the parameters' posterior probability distribution, given the assumptions of large SNR and noise that is stationary and Gaussian \cite{Cutler_Flanagan,Poisson_Will}. 

The probability that the GW data $d(t)$ is characterized by the source parameters $\theta^{a}$ is given by
\begin{equation}\label{posterior}
    p(\boldsymbol{\theta}|d) \propto p^{0}(\boldsymbol{\theta}) \exp\Big[-\frac{1}{2}\Gamma_{ab}(\theta^{a} - \hat{\theta}^{a})(\theta^{b} - \hat{\theta}^{b})\Big],
\end{equation}
where $p^{0}(\boldsymbol{\theta})$ is the prior probability. The values $\hat{\theta}^{a}$ are the maximum of our Gaussian likelihood function and correspond to the ``true'' value of the source parameters in the absence of bias. If the gravitational wave signal is described by $h(t)$, the Fisher information matrix $\Gamma_{ab}$ is given by
\begin{equation}\label{fisher}
    \Gamma_{ab} = \Bigg(\frac{\partial {h}}{\partial \theta^{a}}\Bigg|\frac{\partial {h}}{\partial \theta^{b}}\Bigg),
\end{equation} 
where the inner product is defined as
\begin{equation}
\label{eq:innerprod}
(a|b) = 2 \int_{0}^{\infty}\frac{\Tilde{a}^{\ast}(f) \Tilde{b}(f) + \Tilde{a}(f)\Tilde{b}^{\ast}(f) }{S_{n}(f)}df \,.
\end{equation}
Here $S_{n}(f)$ is the noise spectral density of the detector, $\Tilde{h}(f)$ is the Fourier transform of the time-domain GW signal $h(t)$, and $\ast$ denotes complex conjugation. In practice, the lower and upper limits of integration are defined by the sensitivity of the detector and the source considered.

The prior probability $p^{0}(\boldsymbol{\theta})$ characterizes our prior knowledge about the parameters $\boldsymbol{\theta}$. If $p^{0}(\boldsymbol{\theta})$ follows a Gaussian distribution centered on values $\bar{\boldsymbol{\theta}}$,
\begin{equation}\label{prior}
 p^{0}(\boldsymbol{\theta}) \propto \exp\Big[-\frac{1}{2}\Gamma^{0}_{ab}(\theta^{a} - \Bar{\theta}^{a})(\theta^{b} - \Bar{\theta}^{b})\Big],
\end{equation}
then the covariance matrix is given by
\begin{equation}\label{covariance2}
     \Sigma_{ab} = (\Gamma_{ab}+ \Gamma^{0}_{ab})^{-1},
\end{equation}
where we have assumed $\bar{\boldsymbol{\theta}} \approx \hat{\boldsymbol{\theta}}$.
The 1$\sigma$ statistical errors $\sigma_a$ in the parameters $\theta^{a}$ are given by the square root of the diagonal elements of the covariance matrix:
\begin{equation}\label{error}
    \sigma_{a} = \sqrt{\Sigma_{aa}}\,.
\end{equation}
The parameters of our waveform model are
\begin{equation}\label{parameter space}
    \theta^{a} = \{t_{c}, \phi_{c}, \ln M, \eta, \chi_{1}, \chi_{2}, \ln D_L\}.
\end{equation}
Since the coalescence phase $\phi_{c}$ and spin parameters $(\chi_{1},\chi_{2})$ are restricted to the physically allowed ranges $\phi_{c} \in [-\pi, \pi]$ and $\chi_{1,2} \in [-1, 1]$, we approximately incorporate these constraints in the Fisher matrix approach by adopting Gaussian priors on those parameters with zero means and $1\sigma$ widths given by $\delta \phi_c =\pi$ and $\delta \chi_{1,2}=1$. Note that we work with an angle-averaged waveform, so sky position and binary orientation angles do not enter our analysis; see Appendix \ref{App:waveform} for details.

We use the Cutler and Vallisneri formalism \cite{Cutler_Vallisneri:2007mi} to calculate systematic errors. We can write the approximate waveform in terms of an approximate amplitude ($\mathcal{A}_{\rm AP}$) and approximate phase ($\Psi_{\rm AP}$) via 
\begin{equation}\label{approximate waveform}
    \Tilde{h}_{\rm AP} = \mathcal{A}_{\rm AP}  e^{i \Psi_{\rm AP}}\,.
\end{equation}
The true waveform is similarly written as
\begin{equation}\label{true waveform}
    \Tilde{h}_{\rm T} = \left(\mathcal{A}_{\rm AP} + \Delta\mathcal{A} \right)  e^{i (\Psi_{\rm AP} + \Delta\Psi)} \;, 
\end{equation}
where $\Delta\mathcal{A}$ and $\Delta\Psi$ are the amplitude and phase difference between the true and approximate waveforms, respectively.
The systematic error $\Delta^{\rm sys} \theta^{a}$ in the parameter $\theta^{a}$ is then approximated by \cite{Cutler_Vallisneri:2007mi}
\begin{equation}\label{systematic errors}
    \Delta^{\rm sys} \theta^{a} \approx \Sigma^{ab}\bigg(\big[\Delta\mathcal{A} + i \mathcal{A}_{\rm AP} \Delta\Psi \big] e^{i \Psi_{\rm AP}} \bigg| \partial_{b} \Tilde{h}_{\rm AP}\bigg)\,.
\end{equation} 
(Note that parameter and matrix indices are freely raised or lowered, with repeated indices denoting summation.)
Ignoring the eccentric corrections to the amplitude ($\Delta\mathcal{A}=0$), this equation becomes
\begin{equation}\label{systematic errors_1}
    \Delta^{\rm sys} \theta^{a} \approx \Sigma^{ab}\bigg(i \Tilde{h}_{\rm AP} \Delta\Psi \bigg| \partial_{b} \Tilde{h}_{\rm AP} \bigg) \,,
\end{equation}
where the right-hand side is evaluated at the ``best-fit" value of the parameters. 

In our calculation, the approximate waveform $\Tilde{h}_{\rm AP}$ is the inspiral part of the {\tt IMRPhenomD} waveform, while $\Delta\Psi$ in Eq.~\eqref{systematic errors_1} is the leading-order [i.e., $O(e_0^2)$] eccentric correction to the inspiral stationary phase approximation phasing \cite{Moore:2016qxz} discussed earlier. See \cite{Favata:2021vhw, Saini:2022igm} for more details and applications of systematic errors arising from eccentric corrections. In particular, \cite{Favata:2021vhw} compared the statistical and systematic parameter errors of eccentric binaries using both Bayesian parameter estimation and the Fisher/Cutler-Vallisneri formalism, finding good agreement. That study supports the application of the Fisher/Cutler-Vallisneri formalism in the context of assessing systematic biases in GR tests here and in \cite{Saini:2022igm}.

Note that the eccentric corrections $\Delta \Psi$ that induce systematic parameter errors in Eq.~\eqref{systematic errors_1} are completely decoupled from the {\tt IMRPhenomD} waveform; the latter enters only in $\tilde{h}_{\rm AP}$.  Because of the complexity of the {\tt IMRPhenomD} waveform, the parameter derivatives $\partial \tilde{h}_{\rm AP}/\partial \theta^b$ appearing in Eqs.~\eqref{fisher} and \eqref{systematic errors_1} are not easily computed analytically. We compute them numerically via a ``{symmetric difference quotient}.'' For example, for a function $f(x)$ the derivative is approximated via $f^{\prime}(x) \approx [f(x+h)-f(x-h)]/2h$ for very small $h$.

The noise curve for LIGO is taken from Eq.~(4.7) of \cite{Ajith:2011ec}; the CE noise sensitivity is from Eq.~(3.7) of \cite{Kastha:2018bcr}. The lower cutoff frequency is $f_{\rm low}=10$ Hz for LIGO and $f_{\rm low}=5$ Hz for CE. The transition frequency between the inspiral and merger-ringdown part of the waveform is taken to be the GW frequency $f_{\rm ISCO}$. This frequency corresponds to twice the orbital frequency of a test particle orbiting at the innermost stable circular orbit of the black hole merger remnant formed by the binary coalescence \cite{1972ApJ...178..347B,Husa:2015iqa,Hofmann:2016yih} [see Eq.~(2.23) and Appendix C of \cite{Favata:2021vhw}]. 

Using the {\tt IMRPhenomD} waveform $\tilde{h}(f;\theta^a)$ and the Fisher matrix approach, we compute the $1\sigma$ statistical errors and parameter covariances via three separate approaches: (i) The inspiral-only parameter errors are computed by performing the Fisher matrix frequency-integrals [Eqs.~\eqref{fisher} and \eqref{eq:innerprod}] from $f_{\rm low}$ to $f_{\rm ISCO}$. (ii) The merger-ringdown (MR) parameter errors are obtained by evaluating the Fisher matrix from $f_{\rm ISCO}$ to $f_{\rm end}$. Here $f_{\rm end}$ is chosen to be the minimum frequency beyond which there is no further accumulation of SNR.  (iii) The inspiral-merger-ringdown (IMR) parameter errors involve evaluating the Fisher matrix integrals from $f_{\rm low}$ to $f_{\rm end}$. Note that in all three cases, we are computing the errors in the parameters $\theta^a$ that primarily characterize the {\emph {inspiral}}. Those are the parameters that fully determine the entire {\tt IMRPhenomD} waveform. This is in contrast to the parameters that characterize the BH merger remnant (i.e., via an analysis of ringdown modes), namely the final BH mass $M_f$ and spin $\chi_f$. The computation of the errors in $(M_f, \chi_f)$ is discussed in the next section. The systematic errors in $\theta^a$ [Eq.~\eqref{systematic errors_1}] can only be computed for the inspiral [case (i)], as the eccentric waveform phase correction $\Delta \Psi$ that we employ does not extend to the merger-ringdown. 

\section{Computing errors in the final mass and spin}\label{methods}
Having established a method to compute the statistical and systematic errors for the parameters $\theta^a$, we now proceed to compute the errors in the final mass $M_f$ and final spin parameter $\chi_f$ of the BH remnant. This is effectively done via error propagation as we now describe. 

Defining $\theta^{\mu}=(\ln M, \eta, \chi_1, \chi_2)$ to be a subset of the inspiral parameters $\theta^a$ [Eq.~\eqref{parameter space}], the final mass and spin are related to $\theta^{\mu}$ via numerical relativity (NR) fitting formulas: $M_f=M_f(\theta^{\mu})$ and $\chi_f=\chi_f(\theta^{\mu})$. Here we make use of the NR fits in \cite{Husa:2015iqa,IMRD} (see also Appendix C of ~\cite{Favata:2021vhw} for the relevant formulas). Note that these fits assume that the binary's orbit is circular. In reality, the parameter $e_0$ should be added to $\theta^{\mu}$ as it will affect the relationship between the binary component parameters and the final mass and spin. Here, we assume that the binary already has a modestly low value of $e_0$ ($\lesssim 0.2$) at $f_0$ (near the detector low-frequency cutoff $f_{\rm low}$). As the binary circularizes, the instantaneous eccentricity becomes even smaller near $f_{\rm ISCO}$.\footnote{\label{ftecc} For example, a binary like GW150914 (with masses $m_1=36\, {M}_{\odot}$, $m_2=29\, {M}_{\odot}$ and spins $\chi_1=0.4$, $\chi_2=0.3$) evolves from an eccentricity of $e_0=0.1$ at $10$ Hz to $e\sim0.005$ at the ISCO. A binary with total mass $M=100 M_{\odot}$, mass ratio $q=2$ and spins $(\chi_1,\chi_2)= (0.4,0.3)$ evolves from an eccentricity of $0.1$ at $10$ Hz to $e_t \approx 0.009$ at the ISCO. A system with the same mass ratio and spins but total mass $M=300 M_{\odot}$ (in the CE band) evolves from an eccentricity of $0.01$ at $10$ Hz to $0.003$ at ISCO. We assume $D_L=500$ Mpc for all the systems.}  Hence, we posit that any eccentric correction to the final mass and spin will be small, with $M_f(\theta^{\mu}, e_0)\approx M_f(\theta^{\mu})$ (and likewise for $\chi_f$). We attempt to quantify this approximation in Appendix \ref{NRfitecc} by estimating the eccentric corrections to $M_f$ and $\chi_f$ via a leading-order 0PN (Newtonian) calculation. We find the corrections to be quite small ($\lesssim 1 \%$ for $e_{0} \lesssim 0.2$). NR simulations covering a large parameter space of eccentric binaries will ultimately be necessary to compute these corrections in detail. 

Given the functions $M_f(\theta^{\mu})$ and $\chi_f(\theta^{\mu})$, the corresponding $1\sigma$ statistical errors $(\delta M_f,\delta \chi_f)$ are computed as follows. The Fisher matrix $\Gamma_{ab} + \Gamma^0_{ab}$ determines a Gaussian probability distribution [Eq.~\eqref{posterior}] for all the parameters $\theta^a$ [Eq.~\eqref{parameter space}]. This is then marginalized over the extrinsic parameters $(t_c,\phi_c, \ln D_L)$. In the Fisher matrix approach, this marginalization is performed by simply removing the rows and columns corresponding to $(t_c,\phi_c, \ln D_L)$ in the covariance matrix $\Sigma_{ab}$ [Eq.~\eqref{covariance2}]. This defines a $4 \times 4$ covariance matrix $\Sigma^{(4)}_{\mu \nu}$ for the parameters $\theta^{\mu}$, with diagonal elements defining the squares of the marginalized $1\sigma$ errors $(\delta \ln M, \delta \eta, \delta \chi_1, \delta \chi_2)$. The off-diagonal elements (e.g., $\Sigma^{(4)}_{\ln M\,\chi_1}$) define the marginalized correlations. 

The $1\sigma$ statistical errors and covariances in the final mass and final spin are then computed via standard statistical error propagation using the analytic NR fits $M_f(\theta^{\mu})$ and $\chi_f(\theta^{\mu})$. Defining a $2 \times 2$ symmetric covariance matrix for the parameters $M_f$ and $\chi_f$ as
 \begin{equation}
 \label{eq:SigmaAB}
 \Tilde{\Sigma}_{AB} =
\begin{bmatrix}
\tilde{\Sigma}_{M_f M_f} & \tilde{\Sigma}_{M_f \chi_f} \\
\, & \\
\tilde{\Sigma}_{\chi_f M_f} &  \tilde{\Sigma}_{\chi_f \chi_f}
\end{bmatrix} \,,
\end{equation}
the elements of $\Tilde{\Sigma}_{AB}$ are related to $\Sigma^{(4)}_{\mu \nu}$ via 
\begin{subequations}
\label{eq:SigmaAB2}
\begin{align}
    \tilde{\Sigma}_{M_f M_f} &= \delta M_f^2 = \left(\frac{\partial M_f}{\partial \theta^{\mu}} \right) \left(\frac{\partial M_f}{\partial \theta^{\nu}} \right) \Sigma^{(4)}_{\mu \nu} \,, \\
  \tilde{\Sigma}_{\chi_f \chi_f} &= \delta \chi_f^2 = \left(\frac{\partial \chi_f}{\partial \theta^{\mu}} \right) \left(\frac{\partial \chi_f}{\partial \theta^{\nu}} \right) \Sigma^{(4)}_{\mu \nu} \,, \\
\tilde{\Sigma}_{M_f \chi_f} &= \tilde{\Sigma}_{\chi_f M_f} = \left(\frac{\partial M_f}{\partial \theta^{\mu}} \right) \left(\frac{\partial \chi_f}{\partial \theta^{\nu}} \right) \Sigma^{(4)}_{\mu \nu} \,. 
\end{align}
\end{subequations}
This procedure provides the covariance matrix for the final mass and spin given the covariance matrix of the original parameter set $\theta^a$.

To help interpret and visualize our results in Sec.~\ref{results}, we plot the $1\sigma$ error ellipse in the $M_f$-$\chi_f$ plane. To do this we recognize that the ellipse semimajor $(a)$ and semiminor $(b)$ axes are related to the eigenvalues $(\Lambda_{\pm})$ of the covariance matrix $\tilde{\Sigma}_{AB}$ via\footnote{Note that similar formulas are given in Sec.~VIII of \cite{Favata:2021vhw}, but expressed there in terms of the eigenvalues $\lambda_{\pm}$ and components of the Fisher matrix rather than the covariance matrix. The formulas for $a$, $b$, and $\theta$ are consistent with the results here when using the relation ${\bm \Sigma }={\bm \Gamma}^{-1}$. Note also that a factor of $2$ should multiply the third term on the left side of Eq.~(8.1) in \cite{Favata:2021vhw}.}
\begin{align}
\label{eq:ab}
    a=\sqrt{\Lambda_{+}} \;\;\;\; \text{and}\;\;\;\; b=\sqrt{\Lambda_{-}}\,\,.
\end{align}
The counterclockwise angle $\theta$ of the error ellipse's semimajor axis relative to the $M_f$ (horizontal) axis is
\begin{align}
\label{eq:theta}
    \theta\approx-\frac{1}{2}\arctan\bigg(\frac{2 \tilde{\Sigma}_{M_f\chi_f}}{\tilde{\Sigma}_{\chi_f \chi_f}-\tilde{\Sigma}_{M_f M_f}}\bigg) \,.
\end{align}

We must also propagate the systematic errors $\Delta^{\rm sys} \theta^a$ computed in \eqref{systematic errors_1} to determine the systematic errors in $M_f$ and $\chi_f$ (i.e., $\Delta^{\rm sys} M_{f}$, $\Delta^{\rm sys} \chi_{f}$). In our case, the relations $M_f(\theta^{\mu})$ and $\chi_f(\theta^{\mu})$ provide analytic formulas that relate the shifts $\Delta^{\rm sys} \theta^{\mu}$ (computed for the relevant subset of $\Delta^{\rm sys} \theta^a$) to the shifts $\Delta^{\rm sys} M_{f}$ and $\Delta^{\rm sys} \chi_{f}$. This is simply given by a Taylor series expansion (see, e.g., \cite{eisenhart1963} and Eq.~(2.12) of \cite{ku1966}). For example, the observed (biased) final mass is
\begin{multline}
\label{eq:syserrtaylor}
        M_f(\theta^{\mu}_{\rm obs}) \approx M_f(\theta^{\mu}_{\rm true}) + \frac{\partial M_f}{\partial \theta^{\mu}} (\theta^{\mu}_{\rm obs} - \theta^{\mu}_{\rm true}) \\
        + \frac{1}{2} \frac{\partial^2 M_f}{\partial \theta^{\mu} \partial \theta^{\nu}} (\theta^{\mu}_{\rm obs} - \theta^{\mu}_{\rm true}) (\theta^{\nu}_{\rm obs} - \theta^{\nu}_{\rm true}) + \cdots \,,
\end{multline}
where $\theta^{\mu}_{\rm true}$ are the true values of the parameters, $\theta^{\mu}_{\rm obs}$ are the observed values, and the partial derivatives are evaluated at the true values. Defining $\Delta^{\rm sys} M_f \equiv M_f(\theta^{\mu}_{\rm obs}) - M_f(\theta^{\mu}_{\rm true})$, $\Delta^{\rm sys} \theta^{\mu} \equiv \theta^{\mu}_{\rm obs} - \theta^{\mu}_{\rm true}$, and ignoring the quadratic and higher-order terms, the systematic errors in the final mass and spin are approximated as
\begin{subequations}
\label{eq:sysMchi}
\begin{align}
    \Delta^{\rm sys} M_f &= \frac{\partial M_f}{\partial \theta^{\mu}} \Delta^{\rm sys} \theta^{\mu} \,,  \\
    \Delta^{\rm sys} \chi_f &= \frac{\partial \chi_f}{\partial \theta^{\mu}} \Delta^{\rm sys} \theta^{\mu} \,.
\end{align}
\end{subequations}
We have verified that the quadratic-order corrections to the above change the errors by only $\sim 10\%$. 

\subsection{\label{sec:errorpropDiffvar}Error propagation for the null variables}
In addition to computing the statistical and systematic errors in $M_f$ and $\chi_f$, it is also helpful to define the null variables 
\begin{subequations}
\label{eq:nullvar}
\begin{align}
    \Delta M_f &\equiv M_f^{\rm insp} - M_f^{\rm MR} \,,\\
    \Delta \chi_f &\equiv \chi_f^{\rm insp} - \chi_f^{\rm MR} \,.
\end{align}
\end{subequations}
Here, the variables $\Delta M_f$ and $\Delta \chi_f$ represent the \emph{difference} between the final mass (or spin) computed using only the inspiral signal and only the merger-ringdown (MR) signal. [They are not to be confused with the systematic errors as in Eq.~\eqref{eq:sysMchi}.] In these coordinates the origin point $(\Delta M_f, \Delta \chi_f) = (0,0)$ represents the GR prediction; statistically significant deviations from that origin provide a signature of GR violations via the IMR consistency test. Formulating the IMR consistency test in terms of the null variables provides a simpler interpretation of the overall measurement precision of the test. It also provides a clearer indicator of when systematic errors bias the test to a stated level of statistical significance [e.g., by excluding the $(0,0)$ point].\footnote{To quantify consistency between the inspiral and merger-ringdown, the LVK collaboration's testing GR papers define fractional mass and spin deviation parameters, $\epsilon \equiv \frac{\Delta M_f}{\overline{M}_f} $ and $\xi \equiv \frac{\Delta \chi_f}{\overline{\chi}_f}$. Here $\overline{M}_f$ ($\overline{\chi}_f$) denote the mean of the final mass (final spin) inferred from the inspiral and merger-ringdown. The GR value $(\epsilon,\xi)=(0,0)$ in this parametrization is consistent with the GR value $(0,0)$ in our null parametrization. We use difference (rather than fractional difference) parameters as they greatly simplify the error propagation in our analysis.}

To compute the statistical errors in $(\Delta M_f, \Delta \chi_f)$, we recognize that these null variables are a combination of two separate measurements. In the first measurement, the binary inspiral parameters $\theta^a_{\rm insp}$ [given by Eq.~\eqref{parameter space}] are determined via the {\tt IMRPhenomD} waveform integrated over the inspiral as discussed above (from $f_{\rm low}$ to $f_{\rm ISCO}$). The errors in those variables are determined via the $7\times 7$ dimensional Fisher matrix $\Gamma_{ab}^{(7),\,\rm insp}$ and its corresponding covariance matrix $\Sigma_{ab}^{(7)\,,\rm insp}=\left[\Gamma_{ab}^{(7),\,\rm insp}\right]^{-1}$. One can similarly introduce a new set of variables $\theta^a_{\rm MR}$ that are identical to $\theta^a_{\rm insp}$ except that they represent the values of the inspiral parameters that are determined by applying the {\tt IMRPhenomD} waveform over the merger-ringdown frequency-range only (i.e., from $f_{\rm ISCO}$ to $f_{\rm end}$ as described at the end of Sec.~\ref{sec:formalism}). These variables have errors similarly determined by $7\times 7$ dimensional Fisher and covariance matrices, $\Gamma_{ab}^{(7),\,\rm MR}$ and  $\Sigma_{ab}^{(7)\,,\rm MR}$. 

We can now consider a larger parameter space, $\theta^{\tilde a}=(\theta^a_{\rm insp}, \theta^a_{\rm MR})$, which is the union of the binary parameters measured during the inspiral and the binary parameters measured during the merger-ringdown. The errors in this new parameter set are described by a $14 \times 14$ dimensional Fisher matrix $\Gamma_{\tilde{a}\tilde{b}}^{(14)}$ that is block diagonal and composed of the two seven-dimensional Fisher matrices discussed above:
\be
\Gamma_{{\tilde a}\tilde{b}}^{(14)} =
\begin{bmatrix}
\Gamma_{ab}^{(7),\,\rm insp} & 0^{(7)}_{ab} \\
\, & \\
0^{(7)}_{ab} & \Gamma_{ab}^{(7),\,\rm MR} 
\end{bmatrix}\,,
\ee
where here and below ${\bm 0}^{(n)}$ is a $n\times n$ zero matrix. Because of the block diagonal form, the corresponding covariance matrix can be found by inverting the individual seven-dimensional Fisher matrices:
\begin{align}
\Sigma_{{\tilde a}\tilde{b}}^{(14)} &=
\begin{bmatrix}
\Sigma_{ab}^{(7),\,\rm insp} & 0^{(7)}_{ab} \\
\, & \\
0^{(7)}_{ab} & \Sigma_{ab}^{(7),\,\rm MR} 
\end{bmatrix} \;, \nonumber \\
&= 
\begin{bmatrix}
\left(\Gamma_{ab}^{(7),\,\rm insp} \right)^{-1} & 0^{(7)}_{ab} \\
\, & \\
0^{(7)}_{ab} & \left( \Gamma_{ab}^{(7),\,\rm MR} \right)^{-1} 
\end{bmatrix}\,.
\end{align}
The resulting 14-dimensional covariance matrix can then be marginalized over the parameters $(t_c, \phi_c, \ln D_L)_{\rm insp}$ and $(t_c, \phi_c, \ln D_L)_{\rm MR}$ by removing the corresponding rows and columns. This yields an $8 \times 8$ covariance matrix that is also block diagonal and composed of two $4\times 4$ submatrices:
\be
\label{eq:Sigma8}
\Sigma_{{\tilde \mu}\tilde{\nu}}^{(8)} =
\begin{bmatrix}
\Sigma_{\mu \nu}^{(4),\,\rm insp} & 0^{(4)}_{\mu \nu} \\
\, & \\
0^{(4)}_{\mu \nu} & \Sigma_{\mu \nu }^{(4),\,\rm MR} 
\end{bmatrix} \,.
\ee
Here the rows and columns span the eight-dimensional parameter set $\theta^{\tilde{\mu}} = (\theta^{\mu}_{\rm insp},\theta^{\mu}_{\rm MR})$ consisting of $(\ln M, \eta, \chi_1, \chi_2)$ measured separately during the inspiral and the merger-ringdown. 

Next, following the procedure in Eqs.~\eqref{eq:SigmaAB} and \eqref{eq:SigmaAB2}, we define a $2 \times 2$ covariance matrix that determines the 1$\sigma$ errors and covariances in the null variables,
 \begin{equation}
 \label{eq:SigmaABnull}
 \hat{\Sigma}_{AB} =
\begin{bmatrix}
\hat{\Sigma}_{\Delta M_f \Delta M_f} & \hat{\Sigma}_{\Delta M_f \Delta \chi_f} \\
\, & \\
\hat{\Sigma}_{\Delta \chi_f \Delta M_f } &  \hat{\Sigma}_{\Delta \chi_f \Delta \chi_f}
\end{bmatrix} \,.
\end{equation}
The elements of $\hat{\Sigma}_{AB}$ are related to $\Sigma^{(8)}_{\tilde{\mu} \tilde{\nu}}$ via 
\begin{subequations}
\label{eq:SigmaAB2null}
\begin{align}
    \hat{\Sigma}_{\Delta M_f \Delta M_f} &= \delta (\Delta M_f)^2 = \left(\frac{\partial \Delta M_f}{\partial \theta^{\tilde{\mu}}} \right) \left(\frac{\partial \Delta M_f}{\partial \theta^{\tilde{\nu}}} \right) \Sigma^{(8)}_{\tilde{\mu} \tilde{\nu}} \,,  \\
  \hat{\Sigma}_{\Delta \chi_f \Delta \chi_f} &=  
 \delta (\Delta \chi_f)^2 = \left( \frac{\partial \Delta \chi_f}{\partial \theta^{\tilde{\mu}}} \right) \left(\frac{\partial \Delta \chi_f}{\partial \theta^{\tilde{\nu}}} \right) \Sigma^{(8)}_{\tilde{\mu} \tilde{\nu}} \,,  \\
\hat{\Sigma}_{\Delta M_f \Delta \chi_f} &= 
\hat{\Sigma}_{\Delta \chi_f \Delta M_f} = \left(\frac{\partial \Delta M_f}{\partial \theta^{\tilde{\mu}}} \right) \left(\frac{\partial \Delta \chi_f}{\partial \theta^{\tilde{\nu}}} \right) \Sigma^{(8)}_{\tilde{\mu} \tilde{\nu}} \,.
\end{align}
\end{subequations}
To evaluate the partial derivatives in the above, we note that the functional forms for the NR fits that determine $(M_f,\chi_f)$ depend only on the $\theta^{\mu}$ (i.e., they do not distinguish between $\theta^{\mu}_{\rm insp}$ and $\theta^{\mu}_{\rm MR}$, which represent the same physical parameters determined via different measurement processes). Defining $M_f^{\rm insp}= M_f(\theta^{\mu}_{\rm insp})$, $M_f^{\rm MR}= M_f(\theta^{\mu}_{\rm MR})$, and similarly for $\chi_f^{\rm insp}$ and $\chi_f^{\rm MR}$, the partial derivatives simplify to
\be
\frac{\partial \Delta M_f}{\partial \theta^{\tilde{\mu}}} = \pm \frac{\partial M_f}{\partial \theta^{\mu}} \,,
\ee
where the $(+)$ sign holds for $\theta^{\tilde{\mu}} = \theta^{\mu}_{\rm insp}$, the $(-)$ sign holds for $\theta^{\tilde{\mu}} = \theta^{\mu}_{\rm MR}$, and the right side is evaluated at either $\theta^{\mu} = \theta^{\mu}_{\rm insp}$ or $\theta^{\mu} = \theta^{\mu}_{\rm MR}$. Because of this property and the fact that $\Sigma^{(8)}_{\tilde{\mu} \tilde{\nu}}$ is block diagonal as in Eq.~\eqref{eq:Sigma8}, one can easily show that Eq.~\eqref{eq:SigmaAB2null} simplifies to
\begin{subequations}
\label{eq:SigmaAB2nullB}
\begin{align}
    \hat{\Sigma}_{\Delta M_f \Delta M_f} &= \left(\frac{\partial M_f}{\partial \theta^{\mu}} \right) \left(\frac{\partial M_f}{\partial \theta^{\nu}} \right) \left( \Sigma^{(4), {\rm insp}}_{\mu \nu} + \Sigma^{(4), {\rm MR}}_{\mu \nu} \right) \,,  \\
  \hat{\Sigma}_{\Delta \chi_f \Delta \chi_f} &=  
 \left(\frac{\partial \chi_f}{\partial \theta^{\mu}} \right) \left(\frac{\partial \chi_f}{\partial \theta^{\nu}} \right) \left( \Sigma^{(4), {\rm insp}}_{\mu \nu} + \Sigma^{(4), {\rm MR}}_{\mu \nu} \right)  \,,  \\
\hat{\Sigma}_{\Delta M_f \Delta \chi_f} &= 
\left(\frac{\partial M_f}{\partial \theta^{\mu}} \right) \left(\frac{\partial \chi_f}{\partial \theta^{\nu}} \right) \left( \Sigma^{(4), {\rm insp}}_{\mu \nu} + \Sigma^{(4), {\rm MR}}_{\mu \nu} \right)  \,.
\end{align}
\end{subequations}
The resulting $2\times 2$ covariance matrix can be used to define error ellipses in the $\Delta M_f$-$\Delta \chi_f$ plane via relations analogous to Eqs.~\eqref{eq:ab} and \eqref{eq:theta}.

To compute the systematic error in the null variables we note that
\be
\Delta^{\rm sys}(\Delta M_f) = \Delta^{\rm sys}M_f(\theta^{\mu}_{\rm insp}) \,.
\ee
as there is no eccentricity-induced systematic error in the final mass (and spin) values determined via the merger-ringdown waveform. Hence, the systematic shift in the $1\sigma$ ellipses of the null variables are also given by Eqs.~\eqref{eq:syserrtaylor} and \eqref{eq:sysMchi}:
\begin{subequations}
\label{eq:sysnullvar}
\begin{align}
    \Delta^{\rm sys} (\Delta M_f) &= \frac{\partial M_f}{\partial \theta^{\mu}} \Delta^{\rm sys} \theta^{\mu} \,,  \\
    \Delta^{\rm sys} (\Delta \chi_f) &= \frac{\partial \chi_f}{\partial \theta^{\mu}} \Delta^{\rm sys} \theta^{\mu} \,.
\end{align}
\end{subequations}
\begin{figure*}[pht]
    \centering 
    
  \subfigure{\includegraphics[width=0.49\linewidth]{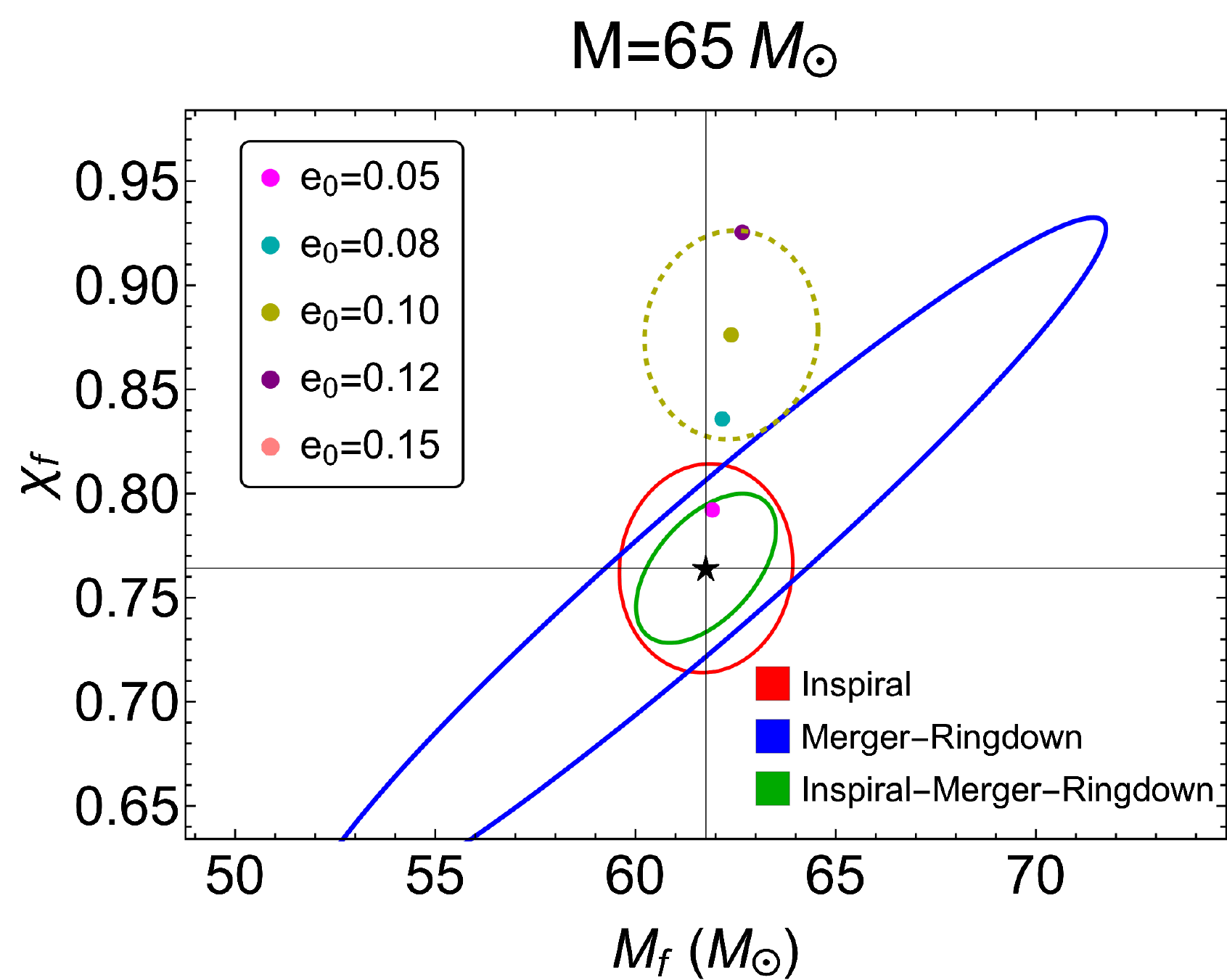}}
  \subfigure{\includegraphics[width=0.49\linewidth]{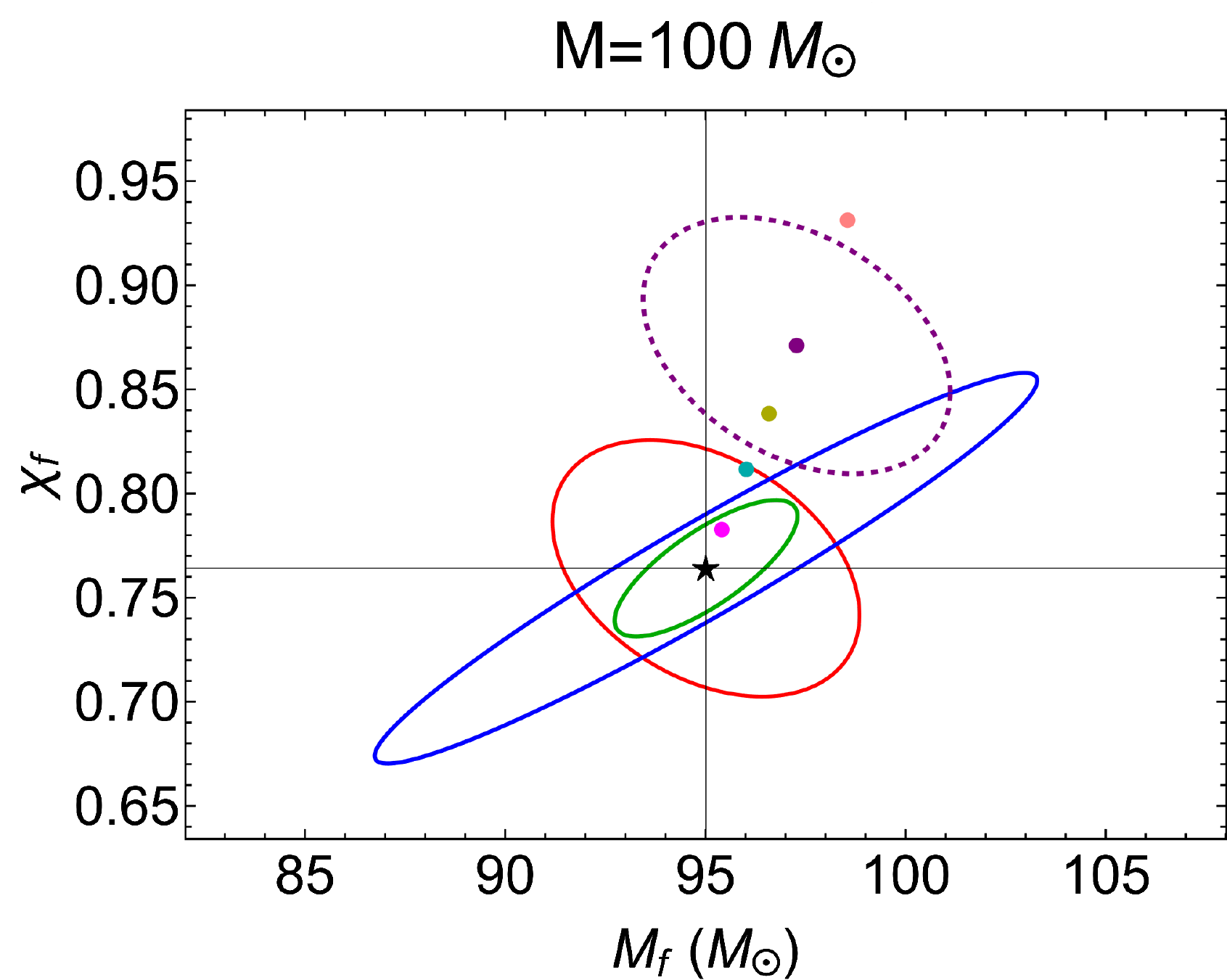}} 
  \subfigure{\includegraphics[width=0.49\linewidth]{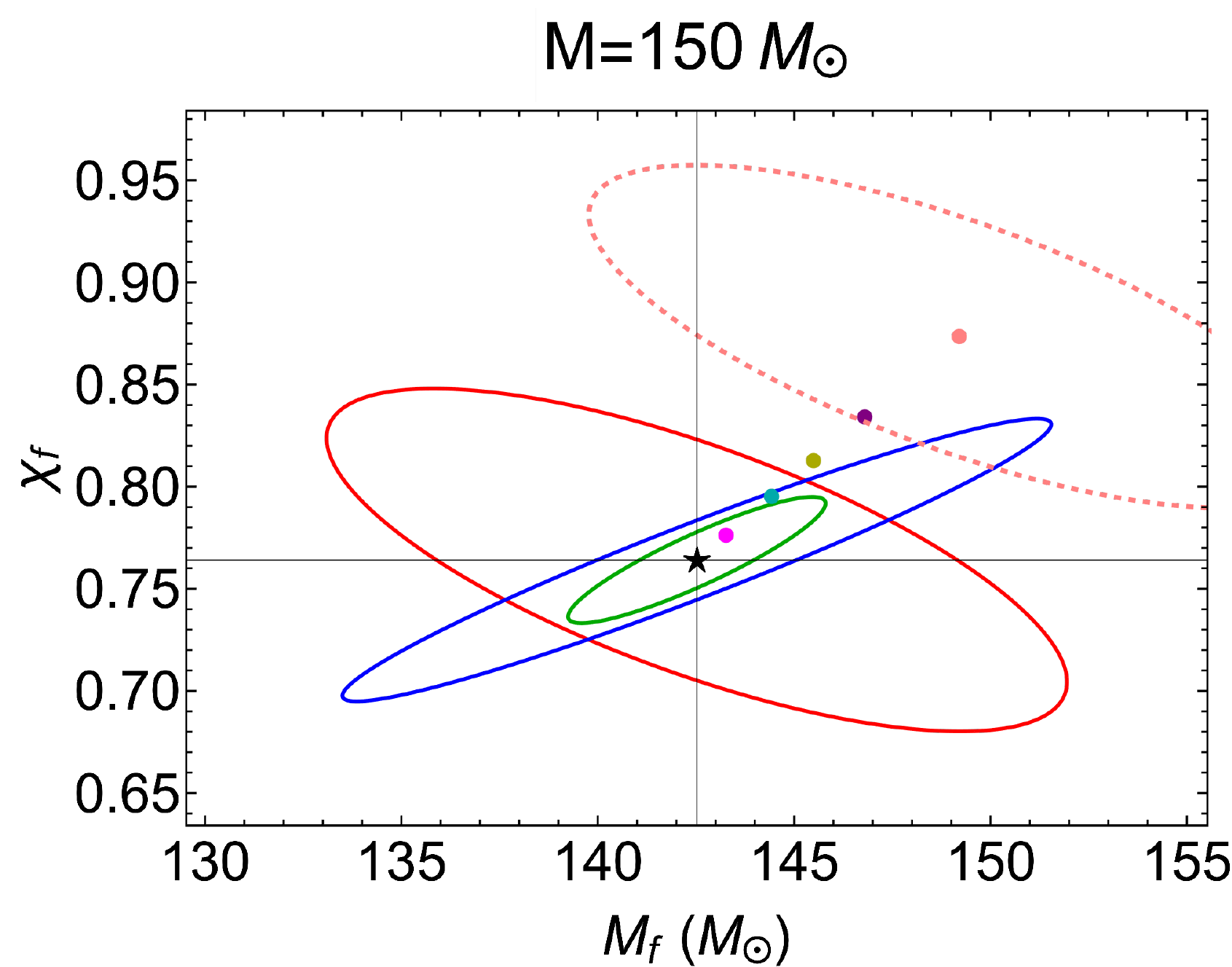}}
   \subfigure{\includegraphics[width=0.49\linewidth]{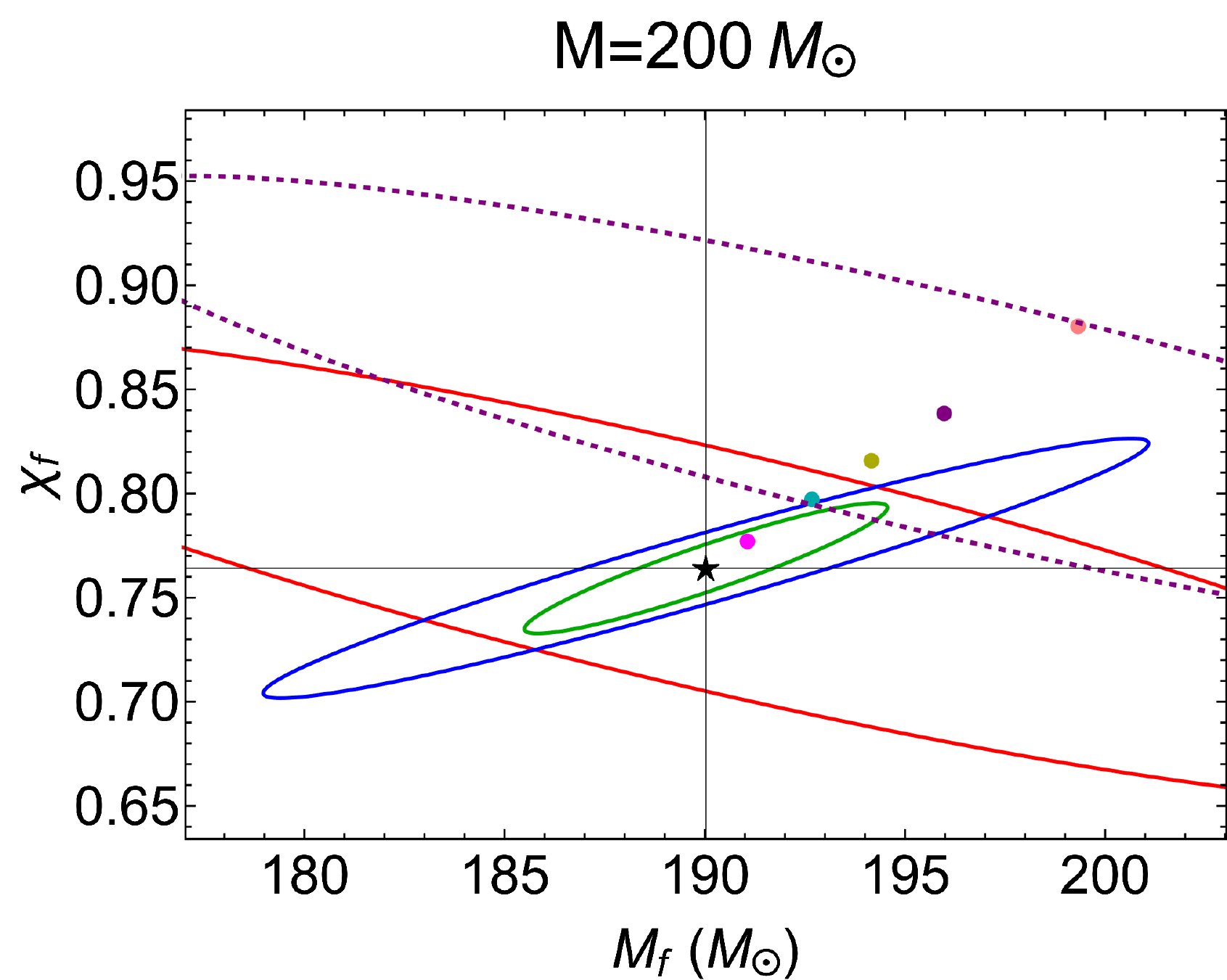}}
  \caption{\label{ellipse_ligo}(Color online) Error ellipses  and systematic bias for selected LIGO band binaries. Ellipses show 1$\sigma$ probability contours in the final mass-final spin plane.  Red ellipses use only information from the inspiral signal, combined with NR fits that determine the final mass and spin. Blue ellipses use signal information from only the merger-ringdown. The green ellipses show the minimum achievable error by using the entire signal waveform (inspiral + merger-ringdown). Ellipse centers (stars) show the final mass and spin values predicted by the NR fits given the binary inspiral parameters and assuming $e_0=0$. Panels are labeled by total binary mass. While the central values differ, each panel shows the same horizontal and vertical scaling for ease of comparison (a range difference of $26 M_{\odot}$ in final mass by $0.35$ in the final spin). Colored points illustrate the systematic bias, showing the migration of the center of the red (inspiral) ellipse as the eccentricity (at $10$ Hz) varies from $e_0=0.0$ (stars) to $0.15$. The dashed curve indicates the biased inspiral ellipse for a single value of $e_0$ (chosen to be the same $e_0$ as the dashed ellipse in the corresponding panel of Fig.~\ref{nullellipse_ligo}). The systematic bias becomes increasingly important as the overlap between the blue (merger-ringdown) ellipse and the shifted inspiral ellipse decreases. For all sources the mass ratio is $2:1$, the luminosity distance is $500$ Mpc, and the dimensionless spin parameters are $\chi_1=0.4$ and $\chi_2=0.3$.} 
\end{figure*} 
\begin{figure*}[pht]
    \centering 
  \subfigure{\includegraphics[width=0.49\linewidth]{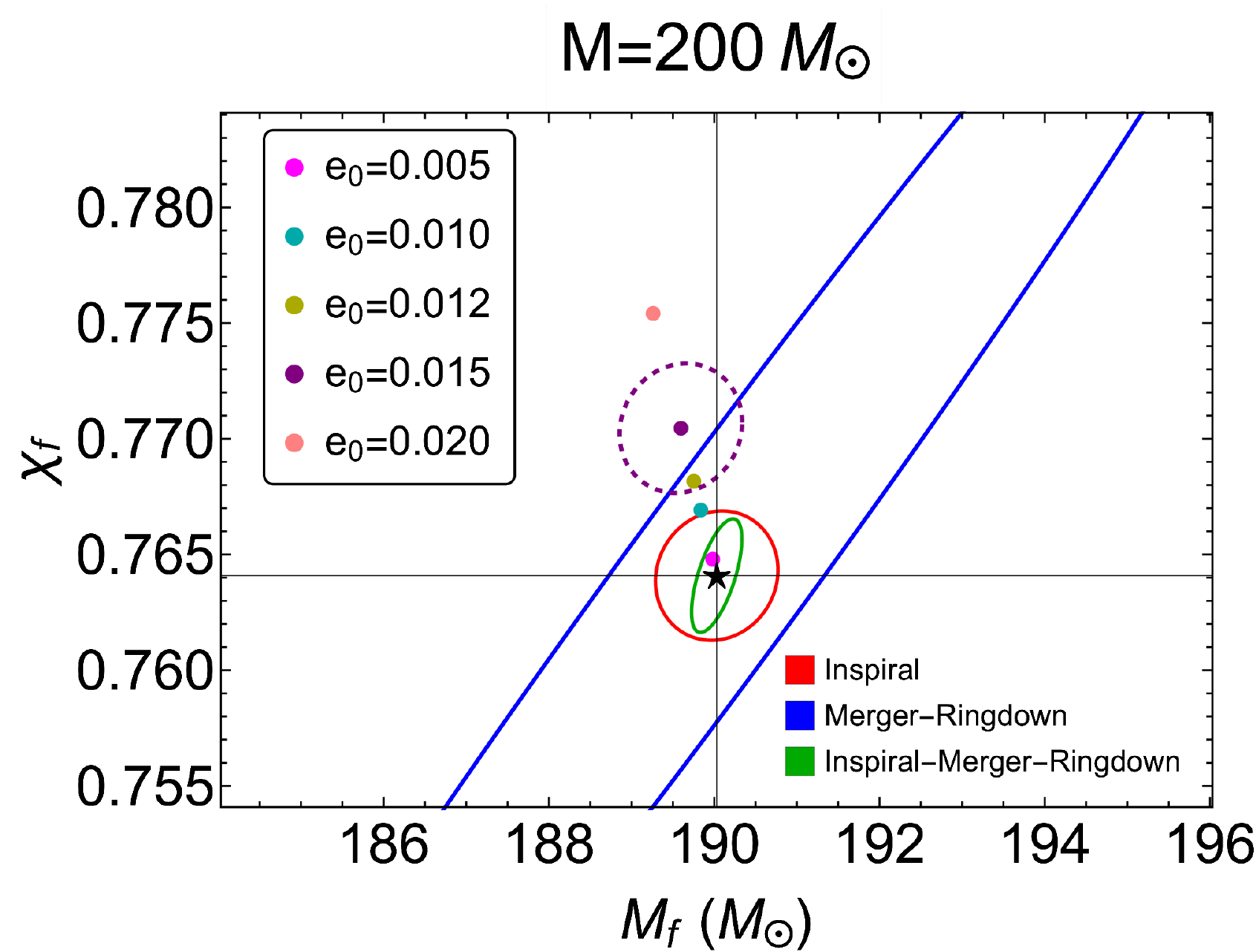}}
  \subfigure{\includegraphics[width=0.49\linewidth]{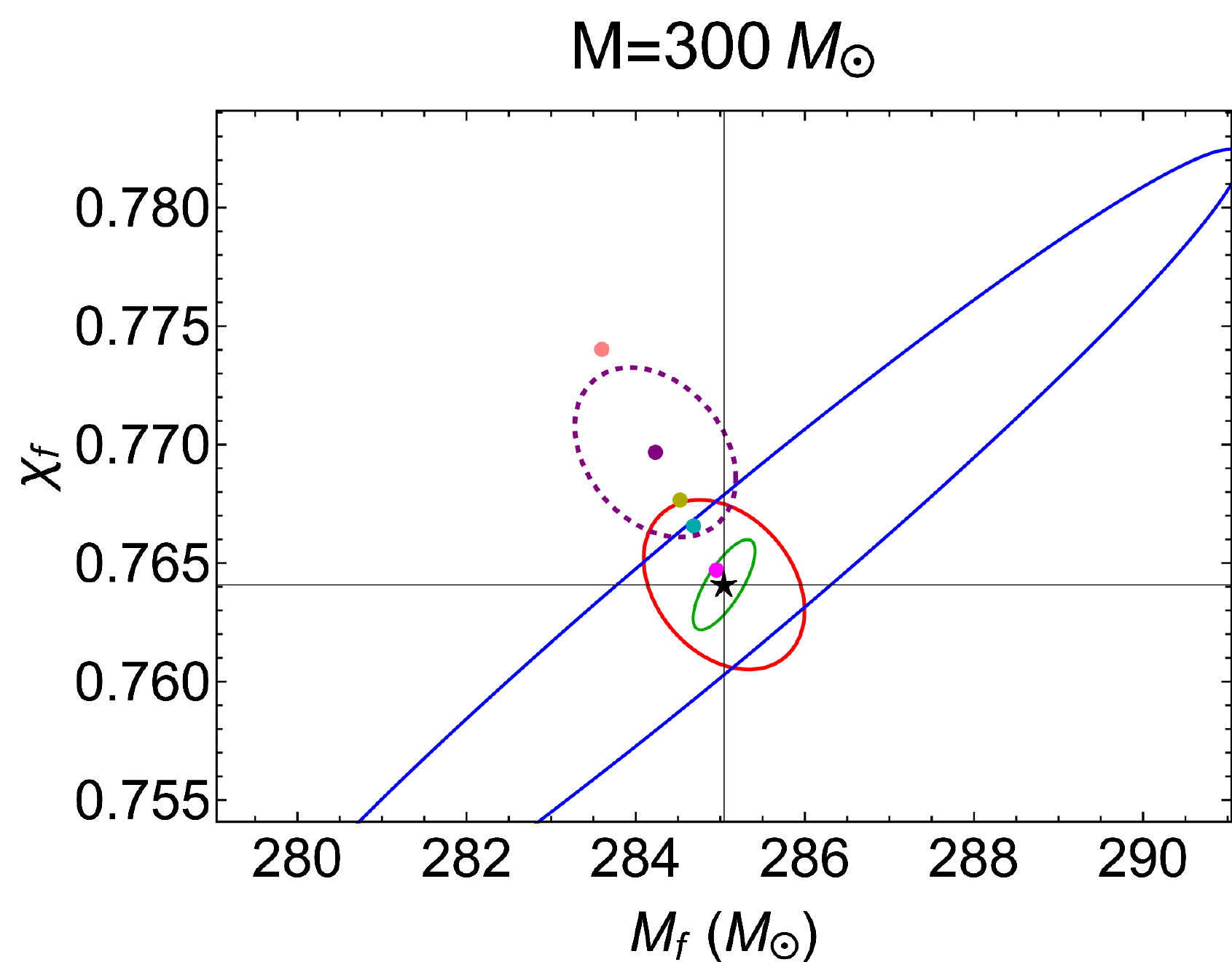}} 
  \subfigure{\includegraphics[width=0.49\linewidth]{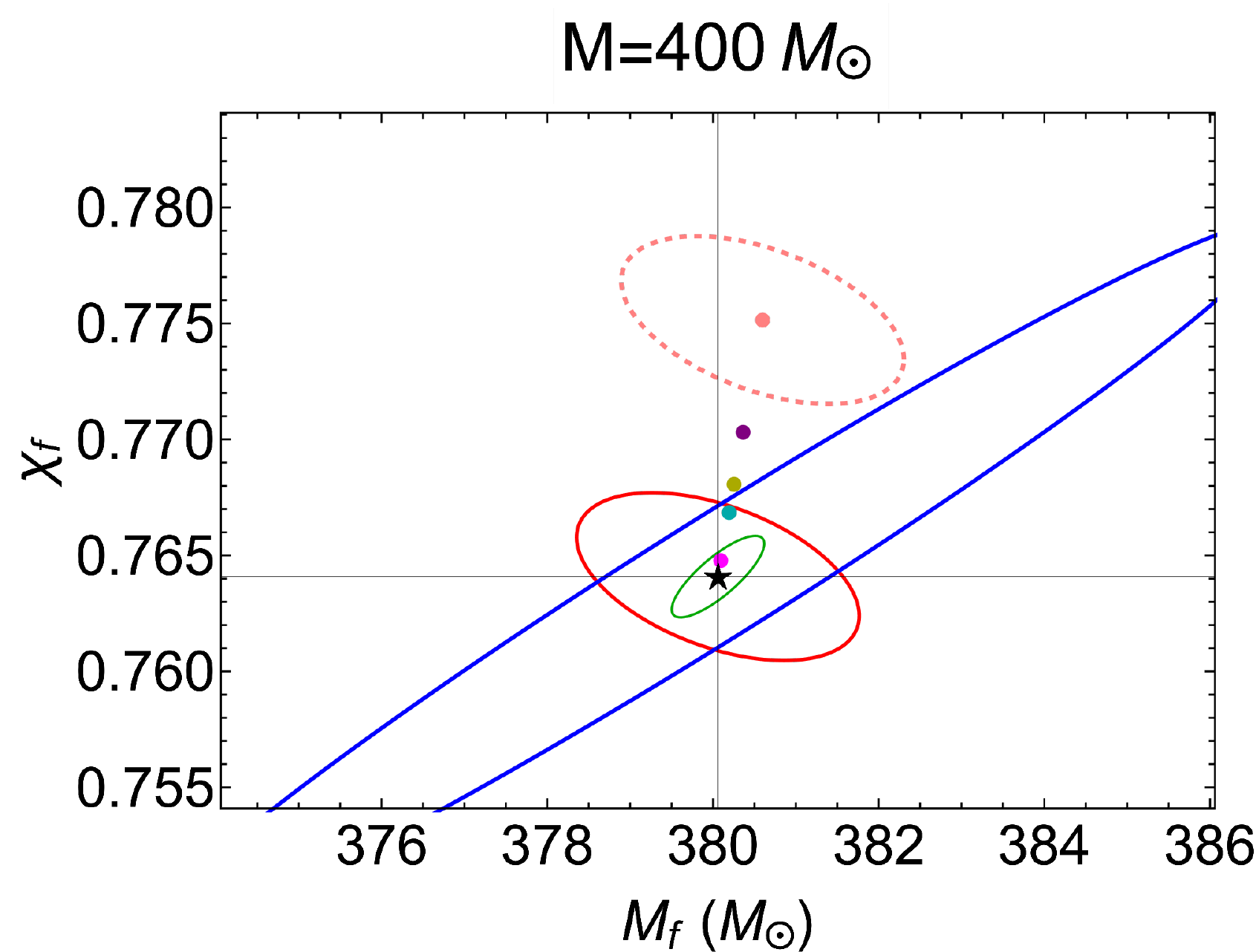}}
   \subfigure{\includegraphics[width=0.49\linewidth]{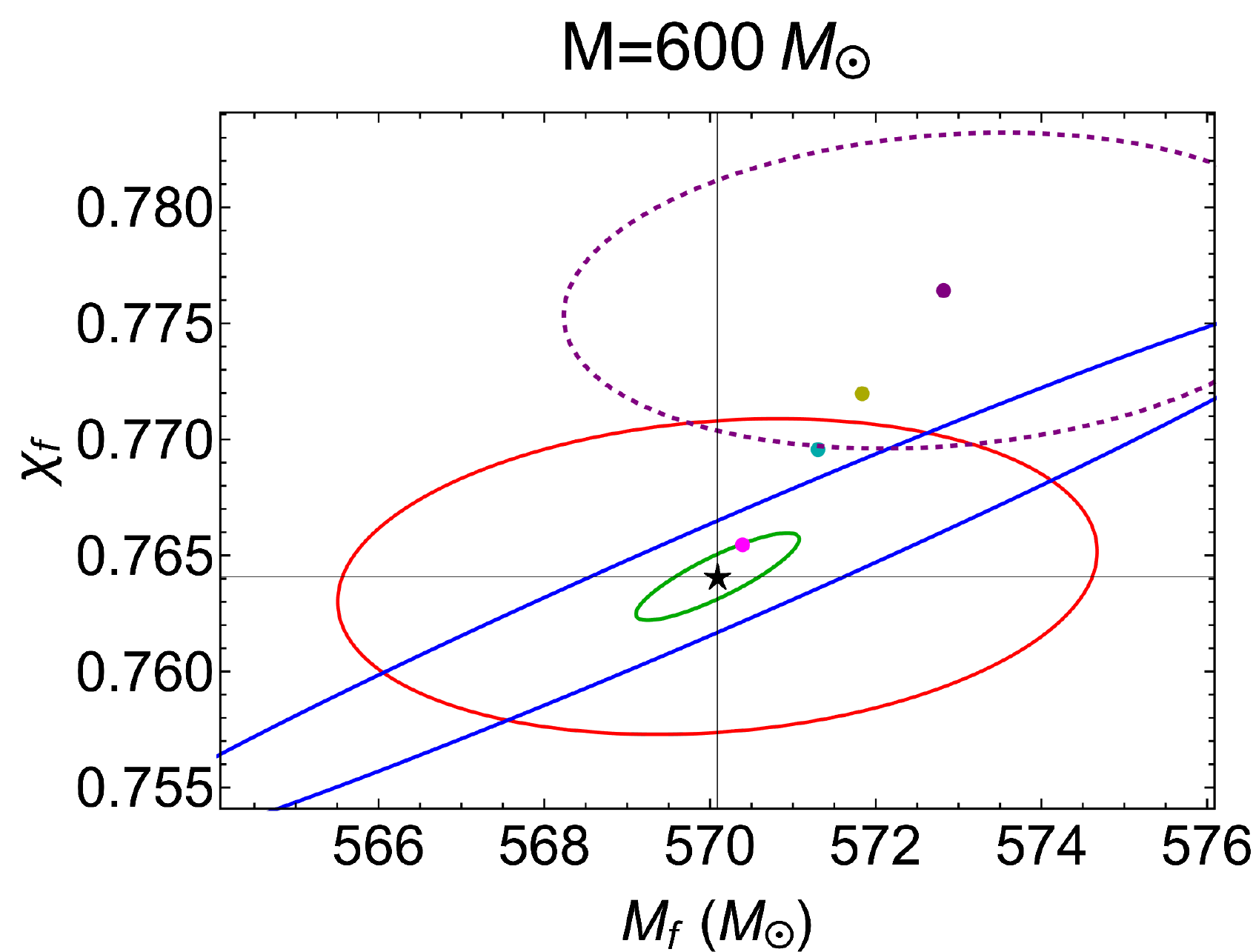}}
   \caption{\label{ellipse_CE}(Color online) Same as Fig.~\ref{ellipse_ligo}, but showing binary black holes in the Cosmic Explorer (CE) band and for a higher range of total masses. All panels are again centered on the final mass and final spin for $e_0=0$, and all have the same horizontal and vertical axis ranges ($12 M_{\odot}$ in final mass by $0.03$ in final spin.) The dashed ellipse is the shifted inspiral ellipse for a particular value of $e_0$ corresponding to the shifted ellipse in Fig.~\ref{nullellipse_CE}.}
\end{figure*} 
\begin{figure*}[pht]
    \centering 
  \subfigure{\includegraphics[width=0.49\linewidth]{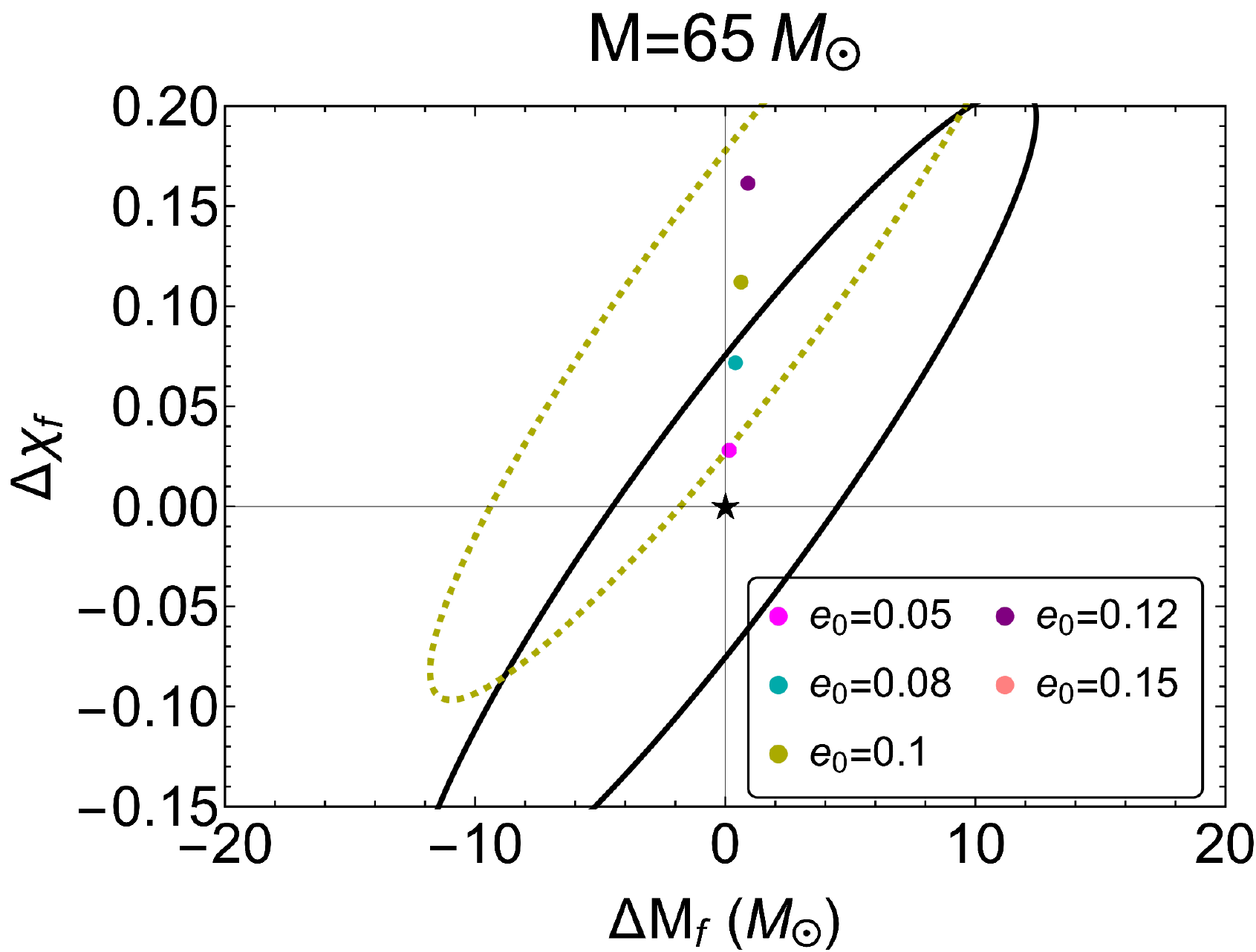}}
  \subfigure{\includegraphics[width=0.49\linewidth]{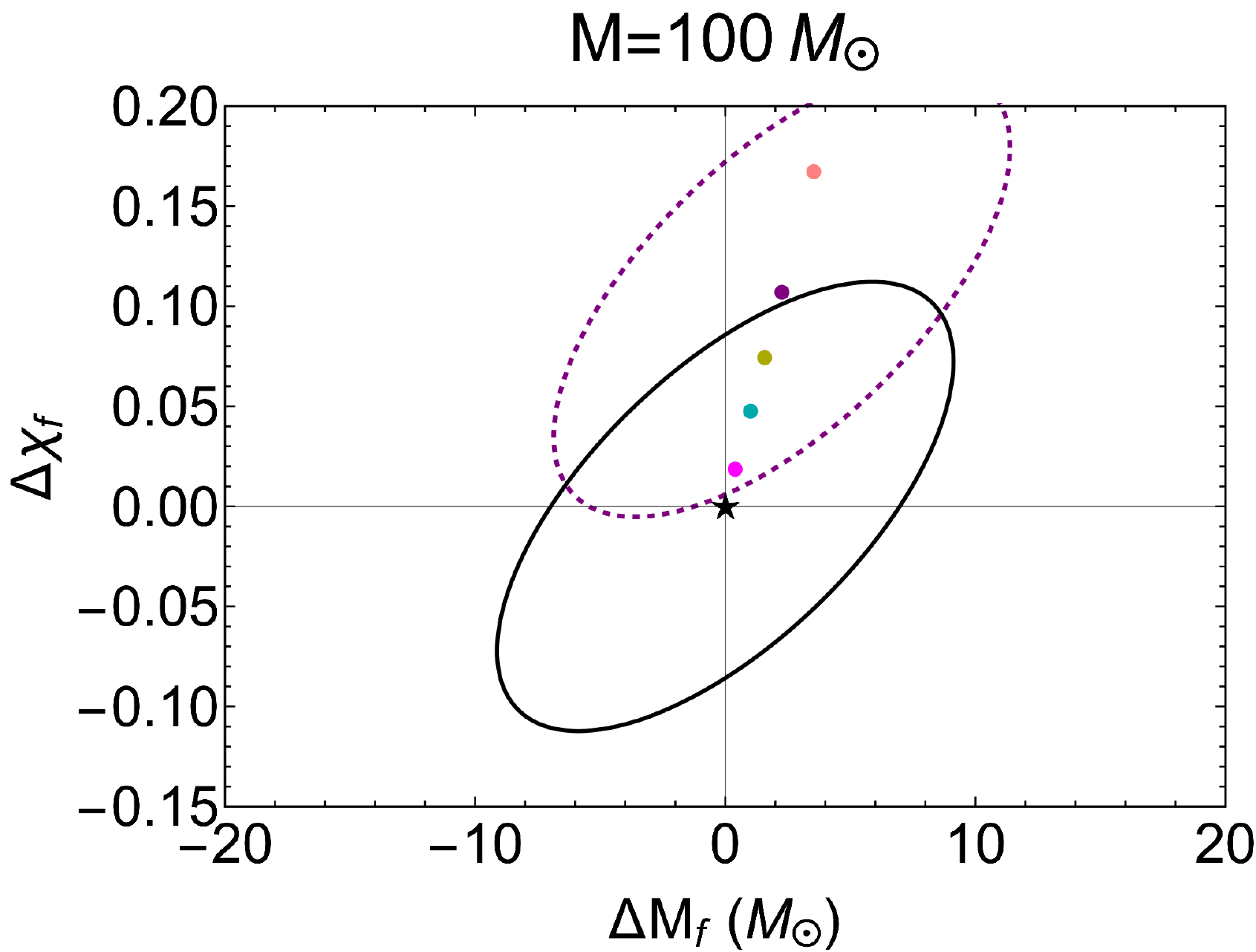}} 
  \subfigure{\includegraphics[width=0.49\linewidth]{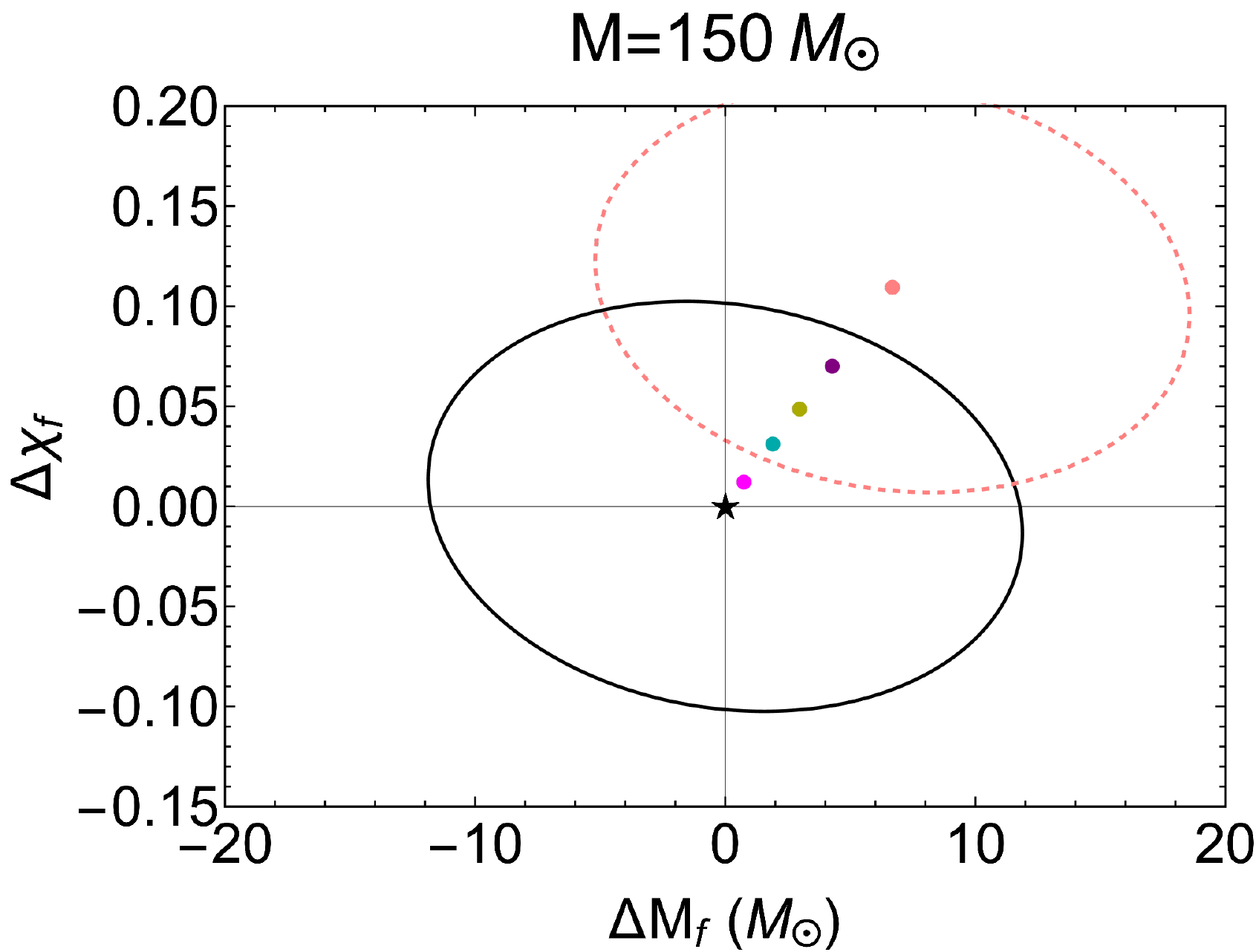}}
   \subfigure{\includegraphics[width=0.49\linewidth]{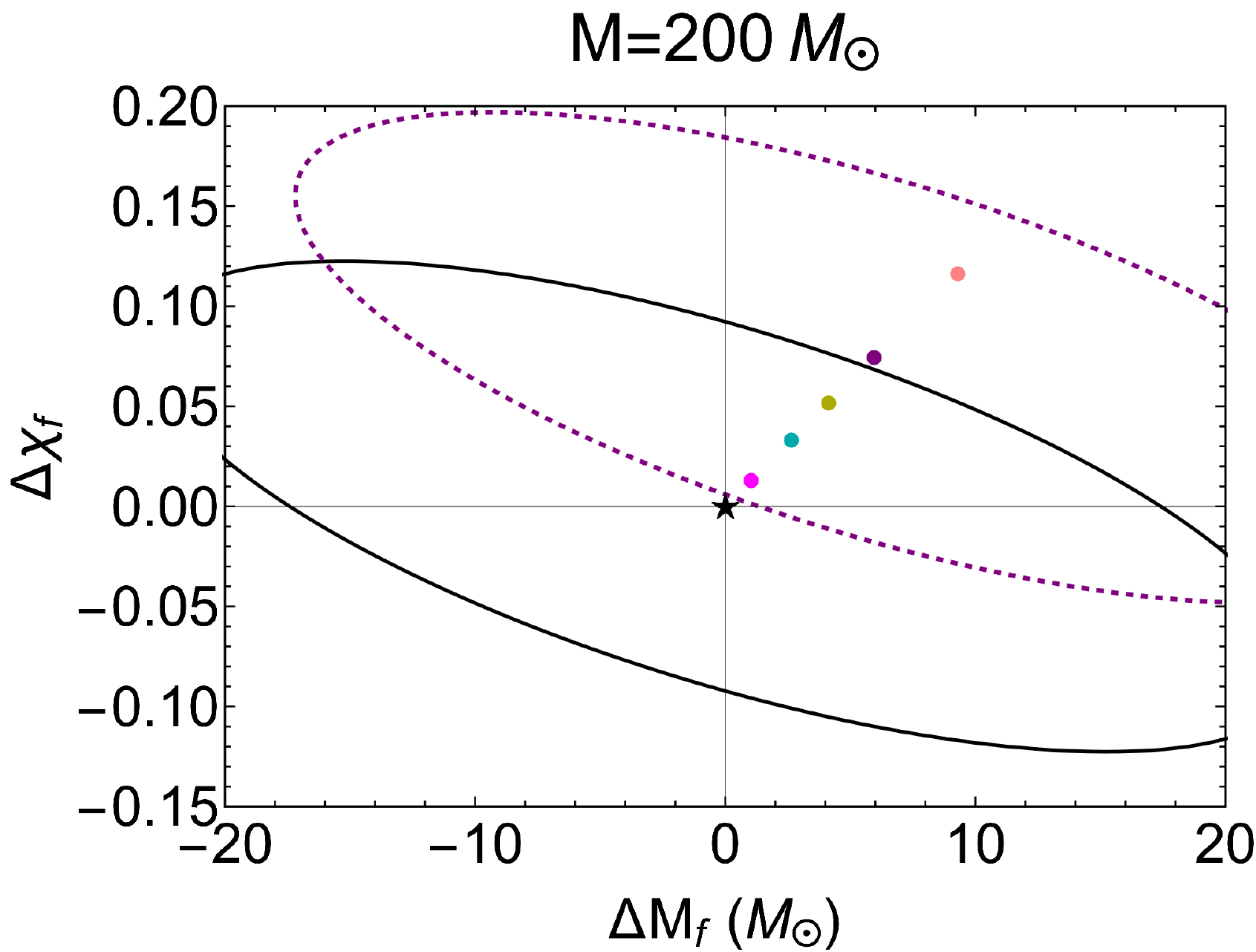}}
    \caption{\label{nullellipse_ligo}(Color online) Error ellipses and systematic bias for selected LIGO band binaries in null variable coordinates. The solid (black) error ellipse shows the two-dimensional measurement precision of the null variables $(\Delta M_f, \Delta \chi_f)$ defined in Eq.~\eqref{eq:nullvar}. These represent the difference between the inferred inspiral and merger-ringdown values of the final mass (or final spin). The central point $(0,0)$ indicates the GR prediction (no difference in the inferred values). Colored dots indicate the shift in the center of the solid (black) ellipse due to the eccentricity-induced systematic bias. The dashed ellipse highlights the shifted ellipse for a particular value $e_0$, as indicated by the central dot of the corresponding color. Note that the highlighted shifted ellipse excludes the origin, indicating an inconsistency with GR at the $1\sigma$ level. The mass and eccentricity values shown here correspond to those in Fig.~\ref{ellipse_ligo}.}
\end{figure*} 
\begin{figure*}[pht]
    \centering 
  \subfigure{\includegraphics[width=0.49\linewidth]{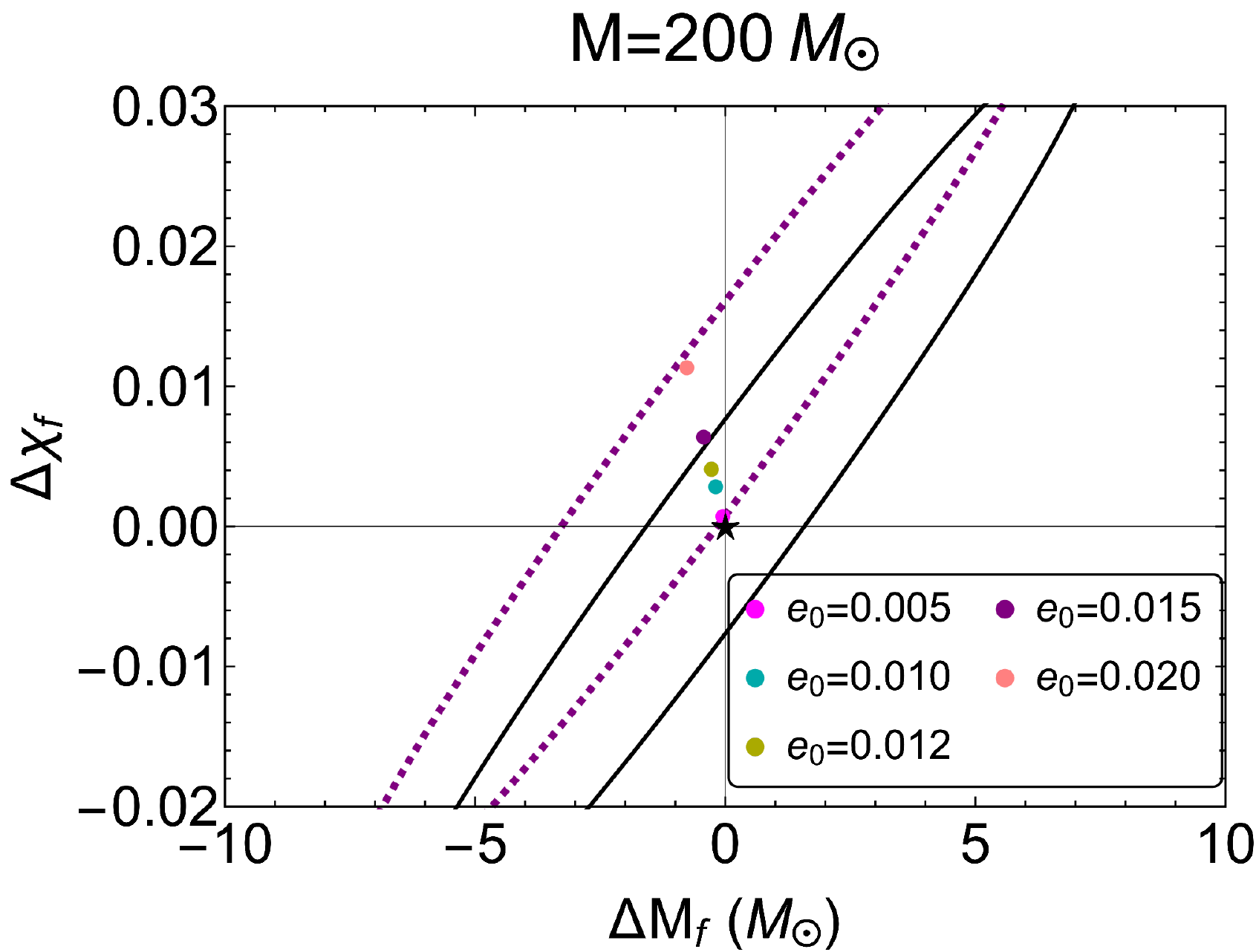}}
  \subfigure{\includegraphics[width=0.49\linewidth]{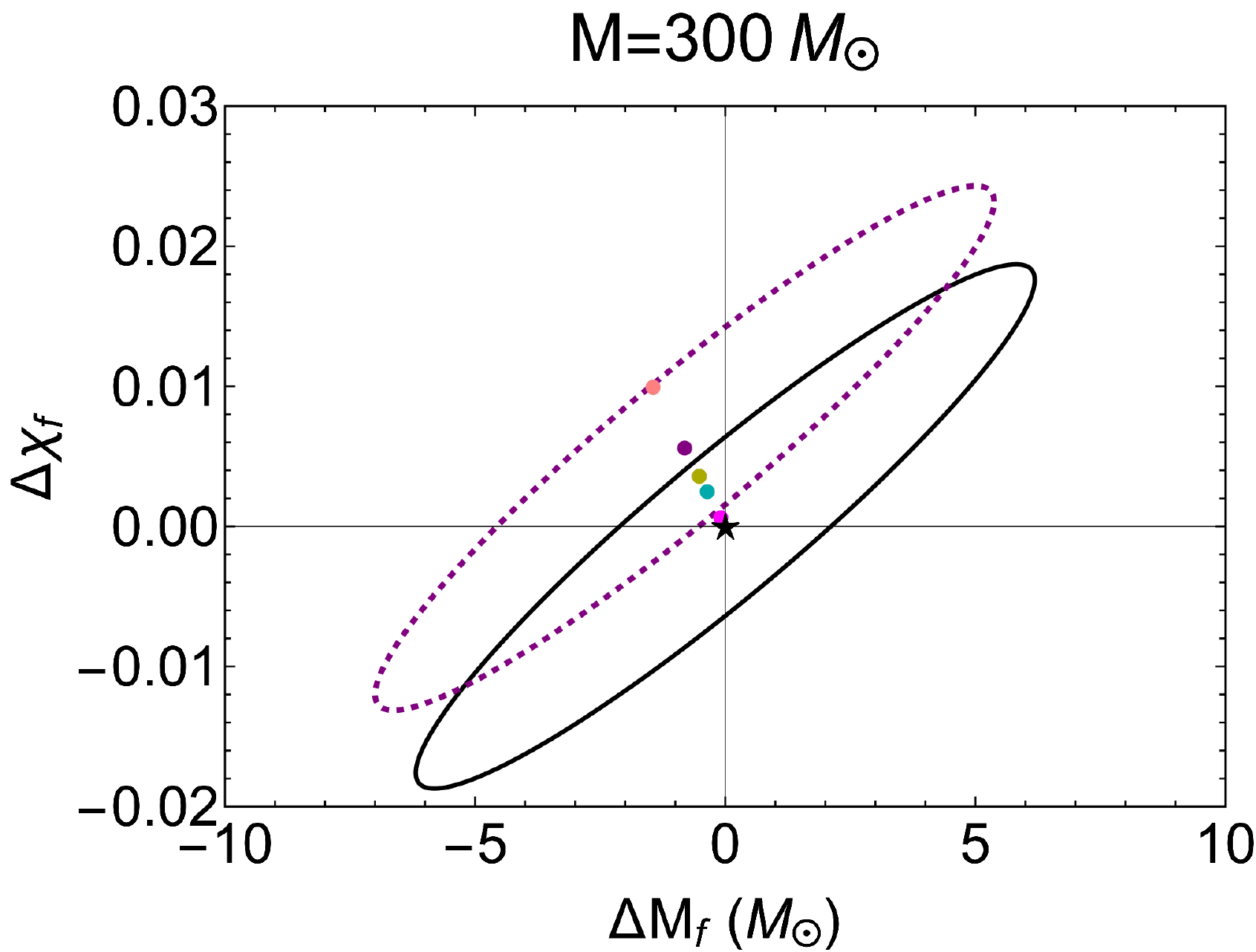}} 
  \subfigure{\includegraphics[width=0.49\linewidth]{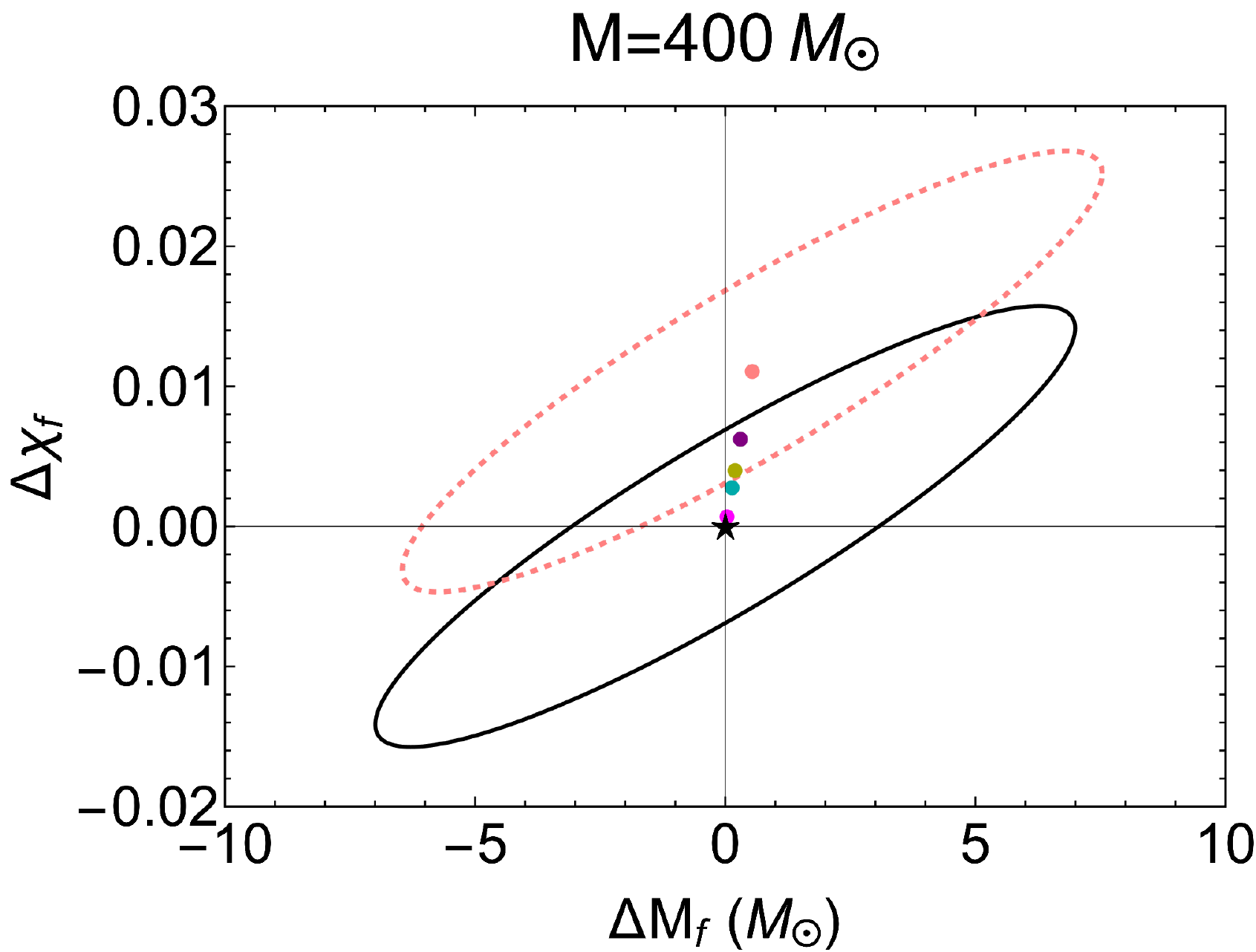}}
   \subfigure{\includegraphics[width=0.49\linewidth]{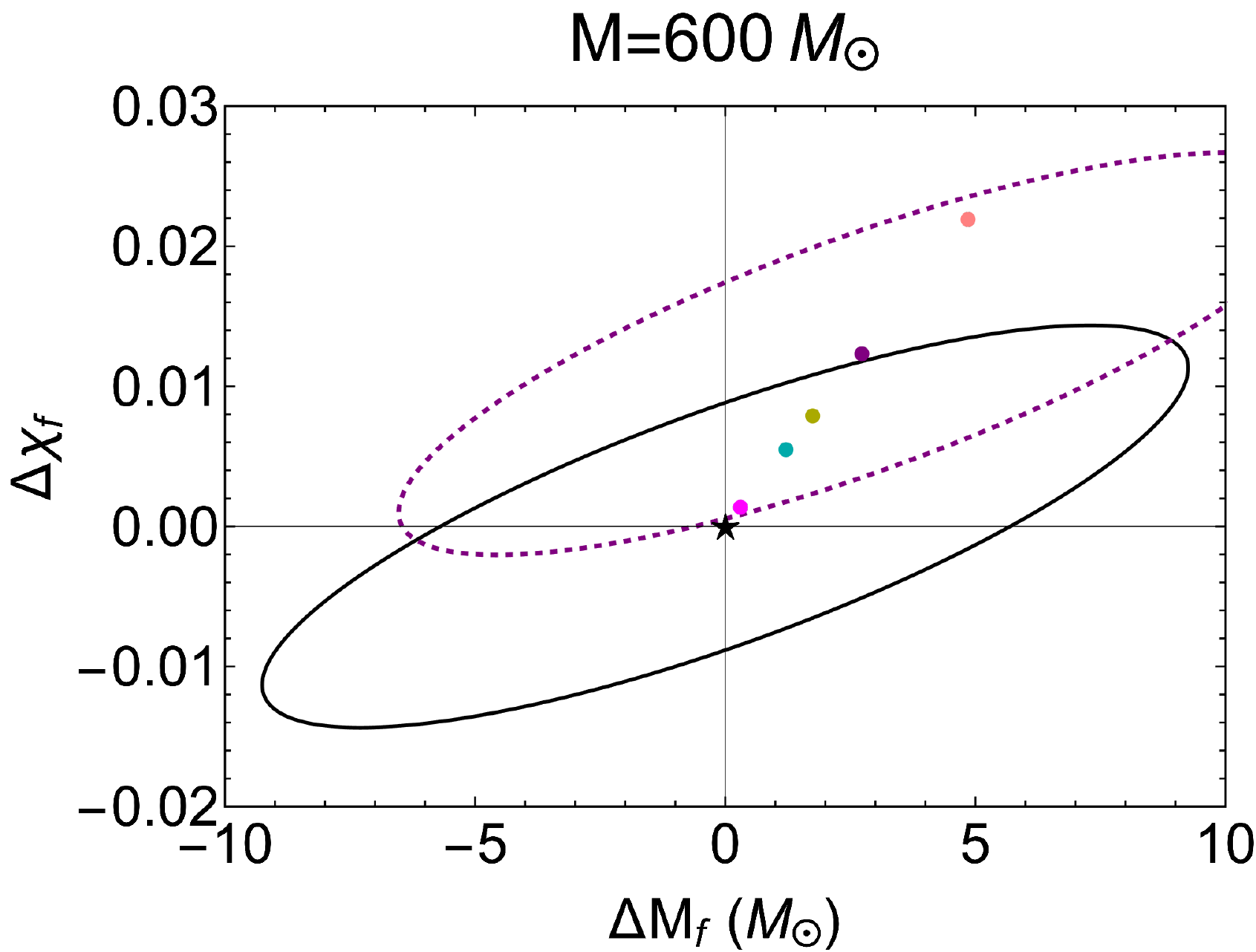}}
  \caption{\label{nullellipse_CE}(Color online) Same as Fig.~\ref{nullellipse_ligo}, but showing the error ellipses and systematic bias for selected CE band binaries in null variable coordinates. The mass and eccentricity values shown here correspond to those in Fig.~\ref{ellipse_CE}. }
\end{figure*} 

\section{Results} \label{results}
Using the formalism described above, we proceed to evaluate the statistical and systematic errors in the final mass and spin, as well as for the corresponding null variables. This is done for binary black hole (BBH) sources in the LIGO and CE bands. We focus on LIGO-band sources with total (source frame) masses of $M=(65, 100, 150, 200)\, M_{\odot}$. For CE (which has better low-frequency sensitivity) we consider more massive binaries with $M=(200, 300, 400, 600)\, M_{\odot}$. In all cases we take the mass ratio to be $m_1/m_2 =2/1$, the BH spins as $\chi_1=0.4$ and $\chi_2=0.3$, and the binary luminosity distance to be $500$ Mpc. 

For LIGO and CE (respectively), Figs. \ref{ellipse_ligo} and \ref{ellipse_CE} show the statistical errors (as ellipses in the $M_f$-$\chi_f$ plane) and systematic errors (as dots). The three error ellipses in each figure panel correspond to the three cases discussed at the end of Sec.~\ref{sec:formalism}: the errors in the final mass and spin inferred via information from the inspiral only (red), the merger-ringdown only (blue), or the entire inspiral, merger, and ringdown (green). In all cases these ellipses are computed by finding the statistical errors in the $\theta^a$ [Eq.~\eqref{covariance2}] and propagating those errors as described by Eqs.~\eqref{eq:SigmaAB} and \eqref{eq:SigmaAB2}. The $1\sigma$ ellipses are drawn via Eqs.~\eqref{eq:ab} and \eqref{eq:theta}. Eccentricity (the source of the systematic error) is assumed to be zero when computing these error ellipses. The figures are centered on the true values of the final mass and spin (represented by a star). The colored dots [computed via Eqs.~\eqref{systematic errors_1} and \eqref{eq:sysMchi} for the inspiral only] show the biased values of the final mass and spin when eccentricity is neglected for eccentric binaries. Each dot represents the biased center of the  \emph{inspiral} error ellipse for a binary with eccentricity $e_0$. 

Considering Fig.~\ref{ellipse_ligo}, we see that the behavior of the statistical errors depends on the system mass. Lower mass systems (e.g., the $65 M_{\odot}$ case) have a long inspiral in the LIGO band but a relatively weak merger-ringdown signal. Hence, the red ellipse is much smaller than the blue one. As the mass increases, the number of inspiral cycles in the LIGO band decreases (leading to larger red ellipses), while the strength of the merger-ringdown signal increases with increasing mass.\footnote{For reference, the SNR as a function of total mass $M= (65, 100, 150, 200)\,M_{\odot}$ is $(32, 42, 51, 55)$ for the inspiral signal and $(13,24,44,62)$ for the merger-ringdown signal.} As the mass increases to $200 M_{\odot}$, the red ellipse expands while the blue ones shrink.  The green error ellipses are the smallest as they combine information from the inspiral and merger-ringdown; they indicate the maximum achievable precision in the final mass and spin if a full waveform model is employed. Note that each panel in Fig.~\ref{ellipse_ligo} shows the same final mass and spin range, so the ellipse sizes can be directly compared. 

Examining the trends in the systematic errors (colored dots), we see a clear trend of increasing deviation from the central value as the eccentricity is increased. The dots represent the migration of the red (inspiral) error ellipse in the $M_f$-$\chi_f$ plane with increasing $e_0$. Systematic errors increase in proportion to $\sim e_0^2$ \cite{Favata:2021vhw}. In the context of the IMR consistency test, one is comparing the final mass and spin inferred from the inspiral (red ellipse) with the final mass and spin inferred from the merger-ringdown (blue ellipse). The dashed ellipse shows the shifted (biased) inspiral ellipse for a particular value of $e_0$. Eccentricity imparts a significant bias on the IMR consistency test when the red (inspiral) error ellipse experiences a sufficient shift such that its overlap with the blue (merger-ringdown) error ellipse is substantially reduced. This happens when $e_0 \gtrsim 0.1$--$0.15$. There are multiple ways to quantify this overlap reduction, and we prefer instead to switch to a null-variable approach, discussed further below.

Figure~\ref{ellipse_CE} similarly shows the error ellipses for sources in the CE band. Because CE is sensitive at lower frequencies than LIGO (down to $\sim 5$ Hz), these binaries have much longer inspirals. Relative to the LIGO band sources, the red (inspiral) error ellipses are therefore much smaller than the blue (merger-ringdown) error ellipses for the cases shown here (up until the $600 M_{\odot}$ case, where the ellipse areas become comparable); this happens for the same reasons as in Fig.~\ref{ellipse_ligo}.\footnote{For reference, the respective SNRs in the CE band for the $M=(200, 300, 400, 600)\, M_{\odot}$ cases are $(2778, 3549, 4059, 4144)$ for the inspiral and $(1122, 2022, 3054, 5188)$ for the merger-ringdown.} For masses below $200 M_{\odot}$, the inspiral error ellipses become significantly smaller than the merger-ringdown error ellipses. We also see that the required value of $e_0$ (which is still defined at a reference frequency of $10$ Hz) to introduce a significant bias is reduced by about a factor of $10$. This is because, for a given value of $e_0$, the instantaneous eccentricity $e_t$ is much greater than $e_0$ over the frequency range $f\in [5,10]$ Hz and hence has a larger impact on the GW phase. In contrast, $f_{\rm low}=f_0$ in the LIGO case, so $e_t<e_0$ over the entire LIGO frequency band. Considering the resulting systematic shift in the inspiral error ellipse, we see significant systematic biases in the IMR consistency test in the CE case for $e_0 \gtrsim 0.015$--$0.020$.

In Fig. \ref{nullellipse_ligo} we plot the measurement precision in the  null-variable plane $(\Delta M_f, \Delta \chi_f)$ via the procedure described in Sec.~\ref{sec:errorpropDiffvar}. The null variable formulation has the advantage of yielding only a single error ellipse that combines the measurement information from both the inspiral and merger-ringdown. Agreement with GR via the IMR consistency test is indicated by an error ellipse centered on the origin in the null-variable plane. Colored dots represent the shift of that ellipse due to the eccentricity-induced systematic bias. If the shifted ellipse excludes the origin, consistency with the GR value $(0,0)$ is excluded with at least $1\sigma$ confidence. The dashed ellipses in Fig. \ref{nullellipse_ligo} highlight selected values of $e_0$ that exclude the origin. This suggests that eccentricity significantly biases the IMR consistency test for $e_0 \gtrsim 0.1$--$0.15$ for LIGO-band binaries. Figure \ref{nullellipse_CE} shows the corresponding ellipses for CE-band binaries. There we see significant biases for $e_0 \gtrsim 0.015$--$0.020$. Note that these values are consistent with the error ellipses showing significant bias in Figs. \ref{ellipse_ligo} and \ref{ellipse_CE}. 

\section{Conclusions}\label{conclusions}
As detections of compact binary coalescences increase, more sensitive tests of GR can be performed in the highly-dynamical and strong-field regime of gravity. But these tests are prone to systematic biases due to any missing physics in the waveform models employed. One might therefore wrongly interpret the resulting systematic error as a GR violation.  Here, we specifically focused on the IMR consistency test: estimates of the BBH remnant's final mass and final spin are computed separately via the inspiral and merger-ringdown signals. Any statistically significant inconsistency in the resulting values could be interpreted as a GR violation.  

We investigate the effect of neglecting orbital eccentricity on the final mass and final spin estimated from the inspiral portion of the signal. Our analysis assumes that eccentricity has a negligible effect on (i) the merger-ringdown signal and (ii) on the relationship between the binary component masses and spins and the BH remnant's final mass and spin. (See Appendix~\ref{NRfitecc} for a justification of the latter.) The resulting bias is investigated for sources in both the LIGO and CE bands. We find that the systematic errors in the remnant final mass and final spin become statistically significant in the CE band if $e_{0} \gtrsim 0.015$ (recall that $e_0$ is defined at a reference frequency of $10$ Hz). In the Advanced LIGO band, systematic errors become statistically significant only at relatively higher eccentricities, $e_{0} \gtrsim 0.1$ at $10$ Hz. 

These results show that neglecting eccentricity in the IMR consistency test might lead to a false claim of a GR violation, even for binary black hole systems with only modest eccentricities. Hence, consistently incorporating eccentricity into waveform models that are applied to these tests is of paramount importance.

\section*{Acknowledgments}
We thank Chandra Kant Mishra for a careful reading of the manuscript.
K.G.A.~acknowledges support from the Department of Science and Technology and Science and Engineering Research Board (SERB) of India via the following grants: Swarnajayanti Fellowship Grant No. DST/SJF/PSA-01/2017-18, Core Research Grant No. CRG/2021/004565, and MATRICS grant (Mathematical Research Impact Centric Support) No. MTR/2020/000177.
K.G.A., S.A.B.~and P.S.~also acknowledge support from the Infosys Foundation. 
M.F.~was supported by NSF (National Science Foundation) Grant No. PHY-1653374. This material is based upon work supported by the LIGO Laboratory, a major facility fully funded by the National Science Foundation.
This paper has been assigned the LIGO Preprint No. P2200216.

\appendix
\section{\label{App:waveform}Relating Fourier amplitudes and phases of IMR waveform modes and polarizations}
Considering only the leading-order $(l,m)=(2,\pm2)$ modes, the plus and cross GW polarizations in the time domain are given by
\begin{align}
{h}_{+} - i {h}_{\times} &={h}_{2,2} \, {}_{(-2)}Y^{2,2}(\Theta,\Phi) + {h}_{2,-2} \, {}_{(-2)}Y^{2,-2}(\Theta,\Phi)  \nonumber \\
&= h_{2,2} \hat{Y}_{+} e^{2 i \Phi} + h_{2,-2} \hat{Y}_{-} e^{-2 i \Phi} \,,
\end{align}
where we used the following spin-weighted spherical harmonic function:
\be
{}_{(-2)}Y_{2,\pm2}(\Theta,\Phi) \equiv \hat{Y}_{\pm}e^{\pm 2 i \Phi} = \alpha (1\pm\cos\Theta)^2 e^{\pm 2 i \Phi} \,,
\ee
with $\alpha = \frac{1}{8} \sqrt{\frac{5}{\pi}}$. Note that
\begin{subequations}
\begin{align}
\hat{Y}_{+} + \hat{Y}_{-} &= 2 \alpha (1+\cos^2\Theta) \;, \\
\hat{Y}_{+} - \hat{Y}_{-} &= 4 \alpha \cos\Theta \;.
\end{align}
\end{subequations}
The angles $(\Theta,\Phi)$ represent the direction to the detector relative to the binary's frame, with $\Theta$ equivalent to the binary inclination angle.  
For nonprecessing binaries orbiting in the $x$-$y$ plane, the $(2,2)$ and $(2,-2)$ modes are related by 
\begin{subequations}
\label{eq:h22relation}
\begin{align}
    h_{2,2}(t) &= h^{\ast}_{2,-2}(t) \;,  \\
    h_{2,2}^{\ast}(t) &= h_{2,-2}(t) \;,
\end{align}
\end{subequations}
where $\ast$ denotes complex conjugation.
We can construct $h_+(t)$ and $h_{\times}(t)$ via
\begin{subequations}
\begin{align}
    h_{+}(t) &= \frac{1}{2} \left[ ( h_{+} - i h_{\times}) + (h_{+} - i h_{\times})^{\ast} \right]  \, \nonumber \\
    &= \frac{1}{2} (\hat{Y}_{+} + \hat{Y}_{-}) \left( h_{2,2} e^{2i\Phi} + h_{2,2}^{\ast} e^{-2i\Phi} \right) \,, \\
    h_{\times}(t) &= -\frac{1}{2i} \left[ (h_{+} - i h_{\times}) - (h_{+} - i h_{\times})^{\ast} \right] \, \nonumber \\
    &=\frac{1}{2} (\hat{Y}_{+} - \hat{Y}_{-}) e^{i\frac{\pi}{2}}  \left( h_{2,2} e^{2i\Phi} - h_{2,2}^{\ast} e^{-2i\Phi} \right) \;.
\end{align}
\end{subequations}
Note that $h_{+,\times}(t)$ are real-valued functions of $t$, while $h_{2,2}(t)$ is a complex function of $t$.

Now we construct the GW strain $h(t) \equiv h_+ F_+ + h_{\times} F_{\times}$ and its Fourier transform. Here $F_{+,\times}(\theta,\varphi, \psi)$ are the detector antenna response functions; they depend on the sky position angles $(\theta,\varphi)$ of the source and the polarization angle $\psi$. From Ref.~\cite{IMRD1} the Fourier transform of the $h_{2,2}(t)$ mode is given by 
\be
\tilde{h}_{2,2}(f) = A_{\rm IMR}(f) e^{-i \phi_{\rm IMR}(f)} \,,
\ee
where $A_{\rm IMR}(f)$ and $\phi_{\rm IMR}(f)$ are real-valued functions that make up the {\tt IMRPhenomD} waveform; they are largely provided in \cite{IMRD1} [see Eqs.~(35), (36), and numerous associated equations in that reference].  This implies that
\be
\tilde{h}_{2,2}^{\ast}(f) = A_{\rm IMR}(f) e^{i \phi_{\rm IMR}(f)} \,.
\ee
The Fourier transform of $h(t)$ can then be written as
\begin{multline}
\tilde{h}(f) = 2 \alpha A_{\rm IMR}(f) \left[ F_{+} (1+\cos^2\Theta)  \cos(\phi_{\rm IMR} - 2\Phi) \right. \\ \left.
 + 2 F_{\times} \cos\Theta  \sin(\phi_{\rm IMR} - 2\Phi)  \right]\,.
\end{multline}
Note that this expression is real valued. Using trig identities we can rewrite this as
\begin{multline}
\tilde{h}(f) = 2 \alpha A_{\rm IMR}(f) \left[ F_{+}^2 (1+\cos^2\Theta)^2 +4 F_{\times}^2 \cos^2\Theta \right]^{1/2} \\ \times \cos(\phi_{\rm IMR} - 2\Phi  - 2\Phi_0) \,,
\end{multline}
where
\be
\Phi_0 = \frac{1}{2} \arctan\left[ \frac{2 F_{\times} \cos\Theta}{F_{+} (1+\cos^2\Theta)} \right] \,.
\ee
Equating to $\tilde{h}(f) = {\rm Re}({\mathcal A} e^{i\Psi})$ we have
\begin{subequations}
\begin{align}
{\mathcal A}(f) &= 2 \alpha A_{\rm IMR}(f) \left[ F_{+}^2 (1+\cos^2\Theta)^2 +4 F_{\times}^2 \cos^2\Theta \right]^{1/2} \,, \\
\Psi(f) &= \phi_{\rm IMR}(f) - 2\Phi  - 2\Phi_0 \,.
\end{align}
\end{subequations}
In our analysis we do not consider the sky position and binary orientation angles as free parameters. 
For an optimally oriented and optimally located binary ($F_{+}=1,\,F_{\times}=0,\Theta=0,\Phi=0$), 
\begin{subequations}
\begin{align}
    {\mathcal A}_{\rm opt}(f) &= 4 \alpha A_{\rm IMR}(f) \,, \\
    \Psi_{\rm opt}(f) &= \phi_{\rm IMR}(f) \,.
\end{align}
\end{subequations}
We can also consider an angle-averaged waveform by averaging $\tilde{h} \tilde{h}^{\ast} = {\mathcal A}^2$ over all the angles $(\Theta, \Phi, \theta, \varphi, \psi)$ (see, e.g., Sec.~IIB of \cite{Favata:2021vhw} where the SNR is also defined). This yields
\begin{subequations}
\begin{align}
    {\mathcal A}_{\rm avg}(f) &= \sqrt{\langle {\mathcal A}^2 \rangle} = \frac{8}{5} \alpha A_{\rm IMR}(f) \,, \\
    \Psi_{\rm avg} &= \phi_{\rm IMR}(f) \,.
\end{align}
\end{subequations}
We use this latter angle-averaged waveform in our analysis. (We note that Ref.~\cite{IMRD1} uses a different convention for the definition of the Fourier transform than what we have used in previous works \cite{Favata:2021vhw}, but this does not affect any of our calculations.) The implementation of {\tt IMRPhenomD} was provided to us by one of the authors via a {\it Mathematica} notebook \cite{KhanNotebook}. It is also available within {\tt LALSuite}\cite{LALSuite}.

\section{\label{NRfitecc}Eccentric correction to the final mass and final spin}
The analysis presented in this paper ultimately relies on the mapping of the binary inspiral parameters $(M,\eta, {\bm \chi}_1,{\bm \chi}_2)$ to the final mass and spin $(M_f,\chi_f)$ of the BH merger remnant. Those relations have been estimated semianalytically for circular orbits in Ref.~\cite{buonanno-kidder-lehner-finalspin}, with accurate relations provided by NR fits \cite{Healy:2016lce,Hofmann:2016yih,Jimenez-Forteza:2016oae}. In a proper generalization of the IMR consistency test, one would need a function $M_f(M,\eta, {\bm \chi}_1, {\bm \chi}_2,e_0)$ that includes the binary eccentricity $e_0$ correction to the final mass (and similarly for the final spin). We have ignored those corrections in our analysis as they are not yet analytically available in the literature. (However, see \cite{Healy:2022wdn} for a recent numerical relativity study that computed the final remnant properties for over 800 eccentric black hole mergers.) Here, we justify this by showing that eccentric corrections to $M_f$ and $\chi_f$ are likely to be very small. Our analysis will rely on a simple quasi-Newtonian analysis of the binary inspiral. It will also assume that the eccentricity is small. 

The final mass is determined by the sum of the binary component masses minus the energy radiated in GWs:
\be
M_f = m_1 +m_2 -\Delta E_{\rm gw} \,,
\ee
where the radiated GW energy is given by minus the change in the orbital energy $E$: $\Delta E_{\rm gw} \approx -\Delta E>0$. This assumes that most of the radiated GW energy is emitted throughout the inspiral up through the last-stable-orbit (LSO) of the binary. (The GW energy emitted during the merger-ringdown is ignored.) The Newtonian orbital energy is
\be
E = -\frac{1}{2} \frac{\mu M}{a} = -\frac{1}{2} \eta M \frac{M}{a} = -\frac{1}{2} \eta M v^2\,,
\ee
where $\mu=m_1 m_2/ M$ is the reduced mass, $M=m_1+m_2$ is the total mass, $\eta=m_1 m_2/M^2$ is the reduced mass ratio, $a$ is the ellipse semimajor axis, and $v\equiv(\pi M f)^{1/3}$. Here we made use of Kepler's third law 
\be
\label{eq:kepler3}
\frac{2}{P} = 2 f_{\rm orb} = f = \frac{1}{\pi} \sqrt{\frac{M}{a^3}}\,, \;\;\;\text{or}\;\;\; \frac{a}{M} = \frac{1}{v^2} \,,
\ee
where $f_{\rm orb}=1/P$ is the orbital frequency, $P$ is the orbit period, and $f$ is the dominant GW frequency if the instantaneous eccentricity $e_t$ is small. Note that these relations are fully valid for eccentric Newtonian orbits. 

Since the orbital energy $E\rightarrow 0$ when the binary is widely separated, the change $\Delta E$ is dominated by the orbital energy at the LSO, $\Delta E=-\frac{1}{2} \eta M v_{\rm LSO}^2$, where $v_{\rm LSO}\equiv (\pi M f_{\rm LSO})^{1/3}$ depends on the GW frequency $f_{\rm LSO}$ at the LSO. The final mass of the BH merger remnant is then given by
\be
\label{eq:Mf}
\frac{M_f}{M} = 1 - \frac{1}{2} \eta v^2_{\rm LSO} \,.
\ee
Note that the eccentricity does not explicitly enter this expression. However, eccentricity does modify the LSO, and this correction is discussed further below.

The final spin vector of the BH remnant is simply the sum of the individual spin vectors of the binary components plus the orbital angular momentum vector at the last stable orbit \cite{buonanno-kidder-lehner-finalspin}, 
\be
{\bm S}_f = {\bm S}_1 + {\bm S}_2 + {\bm L}_{\rm LSO} \,.
\ee
This assumes that the individual spin vectors ${\bm S}_{1,2}$ do not change during the plunge and that little GW angular momentum is radiated during the plunge, merger, and ringdown phases. We will restrict to the case where the spin vectors are aligned or antialigned with the orbital angular momentum vector. The individual BH spins do not precess in this case and we assume they remain constant throughout the inspiral.  We can then replace the above vector equation with an equation for the magnitude of the final spin vector. Using $|{\bm S}|_{1,2} = m_{1,2}^2\, \chi_{1,2}$ gives
\be
\label{eq:chif}
M_f^2 \chi_f = m_1^2 \chi_1 + m_2^2 \chi_2 + L_{\rm LSO}.
\ee
Here $\chi_{1,2} \in (-1,1)$ with positive values indicating spins aligned with the direction of ${\bm L}_{\rm LSO}$. Dividing by $M^2$ gives
\be
\label{eq:chif2}
\left(\frac{M_f}{M}\right)^2 \chi_f = \left(\frac{m_1}{M}\right)^2 \chi_1 + \left(\frac{m_2}{M}\right)^2 \chi_2 + \frac{L_{\rm LSO}}{M^2}.
\ee
Assuming $m_1>m_2$ we can make use of the relation $m_1 M = m_1 (m_1 + m_2) =m_1^2 + \eta M^2$ to show that
\be
\left(\frac{m_{1,2}}{M}\right)^2 = \frac{m_{1,2}}{M} -\eta = \frac{1}{2} \left(1 \pm \sqrt{1-4\eta} \right) -\eta \,,
\ee
where the $+$ in the $\pm$ denotes the $m_1$ case. The magnitude of the orbital angular momentum for an elliptical Newtonian orbit is
\begin{align}
L &= \mu \sqrt{M a (1-e_t^2)} = \mu M \sqrt{(a/M) (1-e_t^2)}  \nonumber\\
&= \frac{\eta M^2}{v} \sqrt{1-e_t^2} \,.
\end{align}
Putting everything together and dividing Eq.~\eqref{eq:chif2} by $(M_f/M)^2$, the final spin becomes
\begin{multline}
\label{eq:chif_final}
\chi_f = \left( 1-\frac{1}{2}\eta v_{\rm LSO}^2 \right)^{-2} \left[ \left( \frac{1}{2} - \eta \right) (\chi_1 + \chi_2) \right. \\
\left. + \sqrt{1-4\eta} (\chi_1-\chi_2) + \frac{\eta}{v_{\rm LSO}} \sqrt{1-e_t(v_{\rm LSO})^2} \right]  \,.
\end{multline}
Here the eccentricity does enter explicitly via the orbital angular momentum. For small eccentricity, $e_t$ varies according to \cite{Moore:2016qxz}
\be
\label{eq:et}
e_t = e_0 \left( \frac{v_0}{v} \right)^{19/6} \,,
\ee
where $v_0\equiv (\pi M f_0)^{1/3}$ for reference frequency $f_0$ (taken to be $10$ Hz). In Eq.~\eqref{eq:chif_final} $e_t$ is evaluated at $v=v_{\rm LSO}$.

The frequency of the LSO (and the corresponding value of $v_{\rm LSO}$) also depends on the eccentricity. As the LSO is not well defined for comparable-mass binaries (eccentric or circular), we will appeal to the extreme-mass ratio limit to estimate the eccentric correction to the LSO. For a point mass orbiting a Schwarzschild BH of mass $M$, the LSO for circular orbits (the ISCO or innermost stable circular orbit) corresponds to $v_{\rm LSO} = v_c \equiv 6^{-1/2}$. To derive a correction in the eccentric case, we use the result in \cite{cutler-kennefick-poissonPRD1994}, which found that eccentric test-mass orbits become unstable for semilatus rectum values $p/M < 6 + 2 e_t$. Equating to the Newtonian definition $p=a (1-e_t^2)$ and using Eq.~\eqref{eq:kepler3}, we arrive at
\begin{align}
v_{\rm LSO} &= 6^{-1/2} \left( \frac{1-e_t^2}{1+\frac{1}{3} e_t} \right)^{1/2} \,, \nonumber \\
&\approx v_c \left[ 1- \frac{1}{6} e_t(v_{\rm LSO}) \right] \,, \nonumber \\
&\approx v_c \left[ 1- \frac{1}{6} e_0 \left( \frac{v_0}{v_{\rm LSO}} \right)^{19/6} \right] \,,
\end{align}
where we expanded in small $e_t$. To solve for $v_{\rm LSO}$ we assume a perturbative solution of the form $v_{\rm LSO} = v_c + \delta v = v_c \left(1 + \frac{\delta v}{v_c} \right)$, where $\delta v \sim O(e_t)$ is small. From this it is clear that an approximate solution is simply
\be
\label{eq:vLSO}
v_{\rm LSO} \approx  v_c \left[ 1- \frac{1}{6} e_t(v_c) \right] \approx  v_c \left[ 1- \frac{1}{6} e_0 \left( \frac{v_0}{v_c} \right)^{19/6} \right] \,.
\ee

With the relevant formulas in hand, we can now evaluate the magnitude of the eccentric correction. First, we note that in the $e_0=0$ case, these quasi-Newtonian formulas already come remarkably close to the values predicted by the NR fits. For example, for an equal mass $(\eta=0.25)$, nonspinning binary $(\chi_1=\chi_2=0)$  and $v_{\rm LSO}=v_c=6^{-1/2}$, we find $M_f = 0.979 M$ and $\chi_f = 0.639$. This agrees with the predicted NR values of $(M_f, \chi_f)= (0.952 M, 0.686)$ to within $2.8\%$ and $6.9\%$, respectively.

To quantify the impact of the eccentric correction we define the fractional error in $M_f$ relative to the circular value $M_f^{(c)} = M_f(e_0=0)$ via
\be
\frac{\delta M_f}{M_{f}^{(c)}} \equiv \frac{M_f - M_f^{(c)}}{M_f^{(c)}} = \frac{\frac{1}{6} \eta v_c^2 e_0 \left( v_0/v_c \right)^{19/6}}{1-\frac{1}{2}\eta v_c^2} \,,
\ee
where we have plugged Eq.~\eqref{eq:vLSO} into Eq.~\eqref{eq:Mf}. 
A similar expression for the final spin is given by
\be
\frac{\delta \chi_f}{\chi_{f}^{(c)}} \equiv \frac{\chi_f - \chi_f^{(c)}}{\chi_f^{(c)}} \approx \frac{1}{6} \left(\frac{5\eta v_c^2-2}{\eta v_c^2-2} \right)  e_0 \left( \frac{v_0}{v_c} \right)^{19/6}\,.
\ee
To arrive at the result above we plugged Eqs.~\eqref{eq:et} and \eqref{eq:vLSO} into \eqref{eq:chif_final}, series expanded in small $e_0$, and assumed $\chi_1=\chi_2=0$ to arrive at a simpler expression. Assuming $\eta=0.25$, $v_c=6^{-1/2}$, and $v_0=(\pi M 10\,{\rm Hz})^{1/3}$, these expressions simply to
\begin{align}
    \frac{\delta M_f}{M_{f}^{(c)}} &= 0.00015 \left(\frac{e_0}{0.1} \right) \left( \frac{M}{100 M_{\odot}} \right)^{19/18} \,,\\
    \frac{\delta \chi_f}{\chi_{f}^{(c)}} &= 0.0032 \left(\frac{e_0}{0.1} \right) \left( \frac{M}{100 M_{\odot}} \right)^{19/18} \,.
\end{align}
For small $e_0$, this is clearly a negligible correction unless $M \gtrsim 1000 M_{\odot}$.

This calculation is approximate and could be improved upon by including higher PN-order terms. Ultimately, NR simulations of merging eccentric binaries will be able to accurately quantify the impact of eccentricity on $M_f$ and $\chi_f$. The analytic study presented here may be helpful in constructing new fitting functions that match those NR results.

\bibliographystyle{apsrev}
\bibliography{ref-list}
\end{document}